\documentclass[journal,12pt,draftclsnofoot,onecolumn,english]{IEEEtran}

\usepackage{epsfig}
\usepackage{times}
\usepackage{float}
\usepackage{afterpage}
\usepackage{amsmath}
\usepackage{amstext,cite}
\usepackage{amssymb,bm}
\usepackage{latexsym}
\usepackage{color}
\usepackage{graphicx}
\usepackage{amsmath}
\usepackage{amsthm}
\usepackage{graphicx}
\usepackage[center]{caption}
\usepackage{pstricks}
\usepackage{caption}
\usepackage{subcaption}
\usepackage{booktabs}
\usepackage{multicol}
\usepackage{lipsum}
\usepackage[T1]{fontenc}
\usepackage{hyperref}
 \usepackage{aecompl}
 \usepackage{mathrsfs}

 \usepackage{arydshln}

\usepackage{soul}
\usepackage{cancel}

\usepackage{changes}
\definechangesauthor[color=red,name={Mingyue Ji}]{MJ}

\allowdisplaybreaks

\long\def\comment#1{}

\newcommand{\bv}{{\mathbf b}}
\newcommand{\cv}{{\mathbf c}}
\newcommand{\dv}{{\mathbf d}}

\newcommand{\fv}{{\mathbf f}}

\newcommand{\sv}{{\mathbf s}}

\newcommand{\Ac}{{\mathcal A}}

\newcommand{\Hc}{{\mathcal H}}

\newcommand{\Kc}{{\mathcal K}}

\newcommand{\Sc}{{\mathcal S}}

\newcommand{\Uc}{{\mathcal U}}

\newcommand{\Zc}{{\mathcal Z}}

\newcommand{\msf}{{\mathsf m}}

\newcommand{\qsf}{{\mathsf q}}

\newcommand{\tsf}{{\mathsf t}}
\newcommand{\usf}{{\mathsf u}}

\newcommand{\vsf}{{\mathsf v}}

\newcommand{\Ksf}{{\mathsf K}}
\newcommand{\Lsf}{{\mathsf L}}
\newcommand{\Msf}{{\mathsf M}}
\newcommand{\Nsf}{{\mathsf N}}

\newcommand{\Rsf}{{\mathsf R}}

\newcommand{\Tsf}{{\mathsf T}}

\newtheorem{thm}{Theorem}

\newtheorem{rem}{Remark}
\newtheorem{example}{Example}

\providecommand{\definitionname}{Definition}

\usepackage{amsmath}
\usepackage{tikz}
\usetikzlibrary{calc}

\usepackage[nodisplayskipstretch]{setspace} 

\newcommand{\tikzmark}[1]{\tikz[overlay,remember picture] \node (#1) {};}
\newcommand{\DrawBox}[4][]{%
    \tikz[overlay,remember picture]{%
        \coordinate (TopLeft)     at ($(#2)+(-0.2em,0.9em)$);
        \coordinate (BottomRight) at ($(#3)+(0.2em,-0.3em)$);
        \path (TopLeft); \pgfgetlastxy{\XCoord}{\IgnoreCoord};
        \path (BottomRight); \pgfgetlastxy{\IgnoreCoord}{\YCoord};
        \coordinate (LabelPoint) at ($(\XCoord,\YCoord)!0.5!(BottomRight)$);
        \draw [red,#1] (TopLeft) rectangle (BottomRight);
        \node [below, #1, fill=none, fill opacity=1] at (LabelPoint) {#4};
    }
}

\begin{document}

\title{On the Tradeoff Between  Computation and Communication Costs  for  Distributed Linearly Separable  Computation}  
\author{
Kai~Wan,~\IEEEmembership{Member,~IEEE,} 
Hua~Sun,~\IEEEmembership{Member,~IEEE,}
Mingyue~Ji,~\IEEEmembership{Member,~IEEE,}  
and~Giuseppe Caire,~\IEEEmembership{Fellow,~IEEE}

\thanks{
K.~Wan and G.~Caire are with the Electrical Engineering and Computer Science Department, Technische Universit\"at Berlin, 10587 Berlin, Germany (e-mail:  kai.wan@tu-berlin.de; caire@tu-berlin.de). The work of K.~Wan and G.~Caire was partially funded by the European Research Council under the ERC Advanced Grant N. 789190, CARENET.}
\thanks{
H.~Sun is with the Department of Electrical Engineering, University of North Texas, Denton, TX 76203, USA (email: hua.sun@unt.edu).
}
\thanks{
M.~Ji is with the Electrical and Computer Engineering Department, University of Utah, Salt Lake City, UT 84112, USA (e-mail: mingyue.ji@utah.edu). The work of M.~Ji was supported in part by NSF Awards 1817154 and 1824558.}
}
\maketitle

\begin{abstract}
 This paper studies the distributed linearly separable computation problem, which is a generalization of many existing distributed computing problems such as    distributed gradient descent and  distributed linear transform. In this problem,
   a master asks $\Nsf$ distributed workers to compute a   linearly separable function of $\Ksf$ datasets, which is  a set of $\Ksf_{\rm c}$ linear combinations of $\Ksf$ messages (each message is a function of one dataset).  We assign some datasets to each worker, which  then computes 
the     corresponding messages and returns some function of these messages, such that   from the answers of any $\Nsf_{\rm r}$  out of $\Nsf$ workers the master can recover the task function. 
 In the literature, the specific case  where   $\Ksf_{\rm c}=1$ or where the   computation cost is minimum has been considered. In this paper, we focus on the general case (i.e., general $K_c $ and general computation cost)  and aim to find  the minimum communication cost.
   
   We first propose a novel converse bound on the communication cost under the constraint   
of the popular {\em cyclic assignment} (widely considered in the literature), which assigns    the datasets  to the workers in a cyclic way.
     Motivated by the observation that existing strategies for distributed computing fall short of achieving the converse bound,  
   we propose a novel distributed computing scheme for  some   system parameters. 
    The proposed computing scheme is   optimal   for any assignment when $\Ksf_{\rm c}$ is large and is optimal under cyclic assignment    when the numbers of workers and datasets are equal or $\Ksf_{\rm c}$ is small. In addition, it is order optimal within a factor of $2$ under cyclic assignment for the remaining cases. 
\end{abstract}


\section{Introduction}
\label{sec:intro}
Nowadays to cope with the emergence of big data and the complexity of data mining algorithm, 
using cloud computing
infrastructures such as Amazon Web Services (AWS) \cite{amazon2015amazon}, Google Cloud Platform \cite{gcp2015}, and 
Microsoft Azure \cite{wilder2012cloud} becomes an efficient and popular solution. 
While large scale distributed computing algorithms and simulations     have the
potential for achieving unprecedented levels of accuracy and
providing dramatic insights into complex phenomena, they
are also presenting new challenges. 
This paper mainly refers to two important challenges of cloud distributed  computing. The first is the relation between the computation and communication
costs. It is critically important to understand the fundamental tradeoff
between computation and communication costs for large scale distributed computing algorithms. 
The second is to tackle the existence of straggler workers (i.e., machines) in applications, such that it is not necessary to wait for the computation of slow workers.
Coding techniques have been introduced into the cloud  distributed computing scenarios~\cite{speedup2018Lee} and have   attracted significant attention recently.
The strategy of this paper is to use coding techniques to characterize the  tradeoff
between computation and communication costs, while mitigating the straggler effect.

  This papers specially considers a  distributed linearly separable computation problem recently formulated in~\cite{linearcomput2020wan}. A master aims to compute a    linearly separable function  $f$  on $\Ksf$ datasets ($D_1,\ldots,D_{\Ksf}$),  where
 $$
 f(D_1,\ldots,D_{\Ksf})= g\big(f_1(D_1), \ldots, f_{\Ksf}(D_{\Ksf}) \big)= g(W_1,\ldots,W_{\Ksf}).
 $$
  $W_k = f_k(D_k)$ for all $k\in \{1, \ldots, \Ksf\}$ is the outcome of the partial function $f_k(\cdot)$ applied to dataset $D_k$.
$g(W_1,\ldots,W_{\Ksf})$ can be seen as a set of  $\Ksf_{\rm c}$    linear combinations of the messages $W_1,\ldots,W_{\Ksf}$ with uniformly i.i.d. coefficients.  
  The task function is computed  by $\Nsf$ workers in the following three phases. 
  During the  {\it data assignment} phase, we assign each dataset to a subset of workers, and the number  of datasets assigned to each worker is defined as the computation cost.\footnote{\label{foot:difference}One of the major differences between this problem and the existing distributed matrix-matrix multiplication problems~\cite{highdimensional2017lee,sparsematrix2018wang,polyoptYu2017,yu2020stragglermitigation,dutta2020optimalRT,struresis2020ramamoo,jia2019matrixCSA}
is that in the considered problem we can only assign the datasets in an uncoded manner to the workers.} 
  During the {\it computing} phase, 
    each worker should compute and send  data packets as functions of the datasets assigned to it, such that 
      from the answers of any $\Nsf_{\rm r}$ 
 workers, the master can recover the task function.  During the {\it decoding}
 phase,   the master should recover the task function by receiving the answers of the $\Nsf_{\rm r}$ fastest  workers. The 
communication cost is defined as the total number of transmissions which should be received by the master in order to recover the task function.
  The objective is to characterize the tradeoff between the computation-communication costs.

In the literature, some sub-cases of the considered problem have been considered. When $\Ksf_{\rm c}=1$, the considered problem becomes   the distributed gradient descent problem considered in~\cite{gradiencoding,MDSGC2018raviv,improvedGC2017halbawi,efficientgradientcoding,adaptiveGC2020}. The optimal   computation-communication costs tradeoff was characterized in~\cite{efficientgradientcoding}  under the constraint of linear coding in  the computing phase and symmetric transmission (i.e., the number of packets transmitted by each worker is the same).
When each worker is limited to  send one linear combination of messages, the considered problem becomes   the distributed linear transform problem treated  in~\cite{shortdot2016dutta}.  The  ``Short-Dot'' distributed computing   scheme was proposed  in~\cite{shortdot2016dutta}, which offers significant speed-up compared to uncoded  computing techniques. When the computation cost is minimum (equal to $\frac{\Ksf}{\Nsf}(\Nsf-\Nsf_{\rm r}+1)$), a distributed computing scheme based on linear space intersection was proposed in~\cite{linearcomput2020wan}, which is exactly optimal when $\Nsf=\Ksf$; and is optimal under the constraint of  cyclic assignment.\footnote{\label{foot:cyclic} The cyclic assignment was  widely used in the existing works on   the sub-problems or related problems of the considered problem such as~\cite{gradiencoding,MDSGC2018raviv,efficientgradientcoding,adaptiveGC2020,linearcomput2020wan,yangelastic2019}. The main advantages of the cyclic assignment are that it can be used for any case where $\Nsf$ divides $\Ksf$  regardless of other system parameters, and its simplicity.  According to our knowledge, the other existing assignments, such as the repetition assignments in~\cite{gradiencoding,replicationcode2020} and the caching-like assignment in~\cite{linearcomput2020wan}, can only be used for very limited number of cases. In addition, the cyclic assignment is independent of the task function; thus if the master has multiple tasks in different times, we need not assign the datasets in each time.
 }

\subsection*{Contributions}
In this paper,  as in~\cite{efficientgradientcoding}, we assume that the computation cost of each worker is $\frac{\Ksf}{\Nsf}(\Nsf-\Nsf_{\rm r}+\msf)$ where $\msf\in [1:\Nsf_{\rm r}]$. Our main contributions are as follows.
\begin{itemize}
\item For any $\msf \in [1:\Nsf_{\rm r}]$,  under the constraint of cyclic assignment, we  propose an information theoretic converse bound on the minimum communication cost $\Rsf^{\star}_{\text{cyc}}$. 
\item On the observation that the existing distributed computing schemes~\cite{efficientgradientcoding,adaptiveGC2020,linearcomput2020wan} for the case $\Ksf_{\rm c}=1$ or $\msf=1$  cannot be used to achieve the converse bound when $\Ksf_{\rm c}>1$ and $\msf>1$, we propose a novel distributed computing scheme under the constraint that $\Nsf \geq \frac{\msf+\usf-1}{\usf}+ \usf(\Nsf_{\rm r}-\msf-\usf+1)$ where $\usf:=\left\lceil \frac{\Ksf_{\rm c} \Nsf}{\Ksf}\right\rceil$. 
\item Compared to the proposed converse bound,  for the considered problem satisfying $\Nsf \geq \frac{\msf+\usf-1}{\usf}+ \usf(\Nsf_{\rm r}-\msf-\usf+1)$,  the proposed computing scheme is exactly optimal when $\Ksf_{\rm c} \in  [\Nsf_{\rm r}-\msf+1: \Ksf]$ and
 is optimal under the constraint of cyclic assignment when $\Ksf=\Nsf$ or $\Ksf_{\rm c} \in \left[1:\frac{\Ksf}{\Nsf} \right]$. In addition, it  is order optimal within a factor of $2$ under the constraint of cyclic assignment for the remaining cases.  
\end{itemize}

 

\subsection*{Paper Organization}
The rest of this paper is organized as follows.
Section~\ref{sec:model} introduces the distributed linearly separable computation  problem and reviews the existing schemes for the case $\Ksf_{\rm c}=1$ or $\msf=1$. 
Section~\ref{sec:main} provides   the main results in this paper and provide some numerical evaluations. Section~\ref{sec:converse} proves the proposed converse bound.
Section~\ref{sec:achie}  describes the proposed   distributed computing scheme. 
Section~\ref{sec:conclusion} concludes the paper and some of the proofs are given in the Appendices.

\subsection*{Notation Convention}
Calligraphic symbols denote sets, 
bold symbols denote vectors and matrices,
and sans-serif symbols denote system parameters.
We use $|\cdot|$ to represent the cardinality of a set or the length of a vector;
$[a:b]:=\left\{ a,a+1,\ldots,b\right\}$   
and $[n] := [1:n]$;   
$a!=a\times (a-1) \times \ldots \times 1$ represents the factorial of $a$;
$\mathbb{F}_{\qsf}$ represents a  finite field with order $\qsf$;         
$\mathbf{M}^{\text{T}}$  and $\mathbf{M}^{-1}$ represent the transpose  and the inverse of matrix $\mathbf{M}$, respectively;
 the matrix $[a;b]$ is written in a Matlab form, representing $[a,b]^{\text{T}}$;
$\text{rank}(\mathbf{M})$ represents the rank of matrix $\mathbf{M}$; 
${\bf 0}_{m \times n}$ represents the zero  matrix with dimension $m\times n$; 
$(\mathbf{M})_{m \times n}$ represents the dimension of matrix $\mathbf{M}$ is $m \times n$;
$\mathbf{M}^{(\Sc)_{\rm r}}$ represents the sub-matrix of $\mathbf{M}$ which is composed of the rows  of $\mathbf{M}$ with indices in $\Sc$ (here $\rm r$ represents `rows'); 
$\mathbf{M}^{(\Sc)_{\rm c}}$ represents the sub-matrix of $\mathbf{M}$ which is composed of the columns  of $\mathbf{M}$ with indices in $\Sc$ (here $\rm c$ represents `columns'); 
 $\text{det}(\mathbf{M})$ represents the determinant matrix $\mathbf{M}$;
 $\text{Mod} (b,a)$ represents the modulo operation with  integer divisor $a$ and in this paper we let $\text{Mod}(b,a)\in \{1,\ldots,a \}$ (i.e., we let $ \text{Mod}(b,a)=a$ if $a$ divides $b$);
we let $\binom{x}{y}=0$ if $x<0$ or $y<0$ or $x<y$.
 In this paper, for each set  of integers  $\Sc$, we sort the elements in $\Sc$ in an increasing order and denote the $i^{\text{th}}$ smallest element by $\Sc(i)$, i.e., $\Sc(1)<\ldots<\Sc(|\Sc|)$.

\section{System Model}
\label{sec:model}

\subsection{Problem formulation}
\label{sub:problem formulation}
We consider a $(\Ksf,\Nsf,\Nsf_{\rm r}, \Ksf_{\rm c}, \msf)$ distributed linearly separable computation problem  over the canonical master-worker distributed system, formulated in~\cite{linearcomput2020wan}.
The master wants to compute a linearly separable function on  $\Ksf$ statistically independent datasets $D_1,  \ldots, D_{\Ksf}$, 
      \begin{subequations}
\begin{align}
   f(D_1, \ldots, D_{\Ksf}) &= g\big(f_1(D_1), \ldots, f_{\Ksf} (D_{\Ksf}) \big) \\
&=  g(W_1, \ldots, W_{\Ksf}),\label{eq:separated objective}
\end{align}
     \end{subequations} 
where we model $f_k(D_k)$,  $k \in [\Ksf]$ as the $k$-th message  $W_k$ and $f_k(\cdot)$ is an arbitrary function. We assume that the $\Ksf$ messages are independent and that each message is composed of $\Lsf$ uniformly i.i.d.   symbols  over a finite field $\mathbb{F}_{\qsf}$  for some large enough prime-power $\qsf$.
As in~\cite{linearcomput2020wan}, we assume that the function $g(\cdot)$ is a  linear mapping as follows,
     \begin{subequations}
\begin{align}
g(W_1,   \ldots, W_{\Ksf})& = {\bf F}  \left[ \begin{array}{c}
W_1\\
\vdots \\
 W_{\Ksf}
\end{array} \right]=
\left[ \begin{array}{c}
F_1\\
\vdots \\
 F_{\Ksf_{\rm c}}
\end{array} \right],
\end{align}
\label{eq:objective matrix-matrix}
     \end{subequations}
where ${\bf F}$ is a matrix known by the master and the workers. The   dimension of ${\bf F}$  is $\Ksf_{\rm c} \times \Ksf$, with elements    uniformly i.i.d.        over  $\mathbb{F}_{\qsf}$. 
The $i^{\text{th}}$ row of ${\bf F}$, denoted by $\fv_i$, is referred to as the $i^{\text{th}}$ demand vector. The $j^{\text{th}}$ element of $\fv_i$  is denoted by $f_{i,j}$.   
It can be seen that  $g(W_1,  \ldots, W_{\Ksf})$ contains 
 $\Ksf_{\rm c} \leq \Ksf$   linear combinations of   the $\Ksf$ messages, whose coefficients are uniformly i.i.d.    over $\mathbb{F}_{\qsf}$.  
 In this paper, 
we assume that  $\frac{\Ksf}{\Nsf} $ is an integer.\footnote{\label{foot:no integer}When $\Nsf$ does not divide $\Ksf$, as shown in~\cite[Section V-A]{linearcomput2020wan}, we can simply add $\left\lceil \frac{\Ksf}{\Nsf}\right\rceil \Nsf-\Ksf $ virtual datasets.}

A distributed computing scheme for our problem contains three phases, {\it data assignment}, {\it computing}, and {\it decoding}. 
\paragraph*{Data assignment phase}
Each dataset $D_k$ where $k \in [\Ksf]$ is assigned to a subset of $\Nsf$ workers in a uncoded manner. 
Define $\Zc_n \subseteq [\Ksf]$ as the set of datasets assigned to worker $n \in [\Nsf]$. 
The   assignment  constraint is that 
\begin{align}
|\Zc_n| \leq \Msf:=\frac{\Ksf}{\Nsf} \left( \Nsf - \Nsf_{\rm r}+\msf \right) , \  \forall  n \in [\Nsf], \label{eq:assignment constraint} 
\end{align}
where  $\Msf:=\frac{\Ksf}{\Nsf} \left( \Nsf - \Nsf_{\rm r}+\msf \right)$ represents the computation cost, and $\msf$ represents the computation cost factor.\footnote{\label{foot:minimum computation}It was proved in~\cite{linearcomput2020wan} that in order to tolerate $\Nsf-\Nsf_{\rm r}$ stragglers, the minimum computation cost is $\frac{\Ksf}{\Nsf} \left( \Nsf - \Nsf_{\rm r}+1 \right)$.}

The assignment function of worker $n$ is denoted by $\varphi_n$, where 
\begin{align}
& \Zc_n = \varphi_n({\bf F}), \\
&\varphi_n  :  [\mathbb{F}_{\qsf}]^{ \Ksf_{\rm c}   \Ksf } \to \Omega_{\Msf}(\Ksf),
\end{align}
and  $\Omega_{\Msf}(\Ksf)$  represents the set of all subsets of $[\Ksf]$ of size not larger than $\Msf$.  
In addition, for each dataset $D_k$ where $k\in [\Ksf]$, we define $\Hc_k$ as the set of workers to whom dataset $D_k$ is assigned.  For each set of datasets $\Kc$ where $\Kc \subseteq [\Ksf]$, we define  $\Hc_{\Kc}:= \cup_{k\in [\Kc]} \Hc_k$ as the set of workers to whom there exists  some dataset in  $\Kc$ assigned.

\paragraph*{Computing phase}
Each worker $n \in[\Nsf]$  first computes the message $W_k = f_k (D_k)$ for each $k \in \Zc_n$.   Worker $n$ then computes 
\begin{align}
 X_n = \psi_n(\{W_k:  k \in \Zc_n\}, {\bf F} )
 \end{align}
  where the encoding function $\psi_n$ is such that
\begin{align} 
\psi_n &:  [\mathbb{F}_{\qsf}]^{ |\Zc_n| \Lsf} \times  [\mathbb{F}_{\qsf}]^{ \Ksf_{\rm c}   \Ksf } \to [\mathbb{F}_{\qsf}]^{ \Tsf_n },  
\label{eq: encoding function def}
\end{align}
and  $ \Tsf_n$ represents the length of $ X_n $. Finally, worker $n$ sends $X_n$ to the master. 

 
\paragraph*{Decoding phase}
The master only waits for   the $\Nsf_{\rm r}$ fastest workers' answers to compute $ g(W_1, \ldots, W_{\Ksf})$. In other words, the computation scheme can tolerate   $\Nsf - \Nsf_{\rm r}$ stragglers.
Since the master does not know a priori which workers are stragglers, the computation scheme should be designed so that from the answers of any $\Nsf_{\rm r}$   workers, the master should recover $g(W_1,   \ldots, W_{\Ksf})$. More precisely, for any subset of workers $\Ac \subseteq [\Nsf]$ where $|\Ac|=\Nsf_{\rm r}$, with the definition
\begin{align}
X_{\Ac}:=\{X_n: n\in \Ac \},
\end{align}
 there exists a decoding function $\phi_{\Ac}$ such that
\begin{align}
&\hat{g}_{\Ac}= \phi_{\Ac}\big( X_{\Ac}, {\bf F} \big)   ,
\end{align}
 where  the decoding function $\phi_{\Ac}$ is such that
\begin{align}
& \phi_{\Ac} :  [\mathbb{F}_{\qsf}]^{\sum_{n \in \Ac} \Tsf_n } \times [\mathbb{F}_{\qsf}]^{\Ksf_{\rm c}   \Ksf}  \to [\mathbb{F}_{\qsf}]^{\Ksf_{\rm c}   \Lsf}.
\end{align}

The  worst-case   probability of
error is defined as
\begin{align}
 \varepsilon:= \max_{\Ac  \subseteq [\Nsf]: |\Ac|= \Nsf_{\rm r}} \Pr\{ \hat{g}_{\Ac} \neq g(W_1,   \ldots, W_{\Ksf}) \}. 
\end{align}

In addition, we denote the communication cost by,  
\begin{align}
\Rsf  :=  \max_{\Ac  \subseteq [\Nsf]: |\Ac|= \Nsf_{\rm r}} \frac{ \sum_{n \in \Ac} \Tsf_n}{   \Lsf  }, \label{eq:communicaton rate}
\end{align}
representing 
 the maximum normalized   number of symbols downloaded by the master from any $\Nsf_{\rm r}$ responding workers.
The communication cost $\Rsf$ is achievable if there exists a computation scheme with assignment, encoding, and decoding functions such that 
\begin{align}
 \lim_{\qsf \to \infty} \  \lim_{\Lsf \to \infty}  \varepsilon =0.
\end{align}
The objective  is to characterize the optimal tradeoff between the computation and communication costs  $(\msf,\Rsf^{\star})$, i.e., for each $\msf \in [\Nsf_{\rm r}]$, we aim to find the minimum  communication cost $\Rsf^{\star}$.

The cyclic assignment was widely used in the  existing works on   the distributed computing problems~\cite{gradiencoding,improvedGC2017halbawi,MDSGC2018raviv,efficientgradientcoding,adaptiveGC2020,linearcomput2020wan}. 
 For each dataset $D_k$ where $k\in  [\Ksf]$, 
we assign $D_k$ to the workers in $\Hc_k$ where (recall that by convention, we let  $\text{Mod}(b,a) =a$ if $a$ divides $b$)
\begin{align}
\Hc_k= \big\{\text{Mod}(k,\Nsf),  \text{Mod}(k-1,\Nsf),\ldots,  \text{Mod}(k-\Nsf+\Nsf_{\rm r}-\msf+1, \Nsf ) \big\}. \label{eq:Hk in cyclic assignment}
\end{align}  
  Thus the set of datasets assigned  to  worker $n \in [\Nsf]$   is 
\begin{align}
\Zc_n=  \underset{p \in \left[0:  \frac{\Ksf}{\Nsf} -1 \right]}{\cup}   \big\{\text{Mod}(n,\Nsf)+ p  \Nsf , \text{Mod}(n+1,\Nsf)+ p  \Nsf , \ldots, \text{Mod}(n+\Nsf-\Nsf_{\rm r}+\msf-1,\Nsf)+ p  \Nsf  \big\}  \label{eq:cyclic assignment n divides k}
\end{align}
with cardinality $\frac{\Ksf}{\Nsf} (\Nsf-\Nsf_{\rm r} +\msf)$.
 For each $\msf \in [\Nsf_{\rm r}]$, the minimum communication cost under  the cyclic assignment in~\eqref{eq:cyclic assignment n divides k} is  denoted   by $\Rsf^{\star}_{\rm cyc}$. 
 
 \begin{rem}
 \label{rem:iid remark}
In the considered problem, the assumption that    the desired function's coefficients (i.e., the coefficients in  demand matrix ${\bf F}$) are uniformly i.i.d.,  
 is needed  to get information theoretic converses and achievability with vanishing probability of error. 
As shown in~\cite[Remark~3]{linearcomput2020wan}, to satisfy  some specific demand matrices,  the optimal communication costs can be  strictly higher
 than $\Rsf^{\star}$.
  It is one of our on-going works to study the arbitrary demand matrices.

 In contrast, the assumption that the symbols in each message are uniformly i.i.d., 
is only needed for the information theoretic converses, while the proposed computing scheme in this paper works for any arbitrary component functions $f_k(D_k)$ where $k\in [\Ksf]$. 
\hfill $\square$ 
 \end{rem}
  
  \subsection{Review of the existing results for $\Ksf_{\rm c}=1$ or $\msf=1$}
\label{sub:review of two cases}
The sub-case  of the considered problem for  $\Ksf_{\rm c}=1$ and any $\msf$ was studied in~\cite{efficientgradientcoding,adaptiveGC2020} and the sub-case for $\msf=1$ and any $\Ksf_{\rm c}$ was studied in~\cite{linearcomput2020wan}. In the following, we review the computing schemes in the literature for the above two sub-cases.

\subsubsection{$\Ksf_{\rm c}=1$}
\label{subsub:review Kc=1}
We first review the computing scheme in~\cite{efficientgradientcoding,adaptiveGC2020} for the case  $\Ksf_{\rm c}=1$.  The cyclic assignment described above  is used for the data assignment phase. In the computing phase, we divide  each message $W_k$, $k\in [\Ksf]$,  into  $\msf$ non-overlapping and equal-length sub-messages  $W_{k}=\{W_{k,i}:i\in [\msf]\}$  where each sub-message contains $\frac{\Lsf}{\msf}$ symbols. Thus the desired linear combination by the master can be seen as $\msf$ linear combinations of sub-messages with the same coefficients.
The main idea is to let each worker send one linear combination of sub-messages, such that the master receives $\Nsf_{\rm r}$ linear combinations of  sub-messages, among which it   then recovers the $\msf$ desired ones. 
We generate $\vsf=\Nsf_{\rm r}-\msf$ virtually demanded linear combinations of sub-messages, such that the effective   demand matrix (containing original and virtual demands) is with dimension $\Nsf_{\rm r} \times \msf\Ksf$ and with the form
\begin{align}
{\bf F^{\prime}} =\begin{bmatrix}       
f_{1,1} &\cdots &f_{1,\Ksf}& 0 &\cdots  & 0& \cdots & 0 & \cdots &0 \\
0 & \cdots & 0 & f_{1,1} & \cdots & f_{1, \Ksf} & \cdots&  0& \cdots &0 \\
\vdots &\ddots & \vdots &\vdots  &\ddots  & \vdots & \ddots& \vdots &\ddots &\vdots \\
0 & \cdots & 0 &0 & \cdots & 0 & \cdots& f_{1,1}& \cdots &f_{1,\Ksf} \\ 
  a_{1,1}&\cdots  &   a_{1,\Ksf} & a_{1,\Ksf+1} &\cdots & a_{1,2\Ksf}& \cdots & a_{1,(\msf-1)\Ksf+1} &\cdots  & a_{1, \msf \Ksf} \\
\vdots &\ddots & \vdots &\vdots  &\ddots  & \vdots & \ddots& \vdots &\ddots &\vdots \\
      a_{\vsf,1}&\cdots  &   a_{\vsf,\Ksf} & a_{\vsf,\Ksf+1} &\cdots & a_{\vsf,2\Ksf}& \cdots & a_{\vsf,(\msf-1)\Ksf+1} &\cdots  & a_{\vsf, \msf \Ksf} 
\end{bmatrix}. \label{eq:Kc=1 F prime}
\end{align}
 The transmission of worker $n \in[\Nsf]$ can be expressed as 
 $$
\sv^{n,1} \ {\bf F^{\prime}}  \ [W_{1,1};W_{2,1};\ldots; W_{\Ksf,1};W_{1,2};\ldots;W_{\Ksf,\msf}],
$$    
 where $\sv^{n,1}=(s^{n,1}_1,\ldots,s^{n,1}_{\Nsf_{\rm r}} )$ is the transmission vector for worker $n$. The next step is to determin the values for each $\sv^{n,1}$ where $n\in [\Nsf]$. The authors in~\cite{efficientgradientcoding} choose these values from 
    a specific matrix while the authors in~\cite{adaptiveGC2020} choose the value of each element in these vectors uniformly i.i.d over $\mathbb{F}_{\qsf}$. Here  we use the random generation    in~\cite{adaptiveGC2020}.  
    Let us then focus on each column in ${\bf F^{\prime}} $, which corresponds   to a sub-message. For example, the first column of ${\bf F^{\prime}} $ corresponds to $W_{1,1}$, which   
     cannot be computed by  $     \Nsf_{\rm r}-\msf=\vsf$ workers, i.e., the workers in $[\Ksf]\setminus \Hc_1$. Hence, for each worker $n \in ([\Ksf]\setminus \Hc_1)$, it should satisfy
     \begin{align}
    0&= s^{n,1}_1 f_{1,1} + s^{n,1}_2 0 +\ldots + s^{n,1}_{\msf} 0 +  s^{n,1}_{\msf+1}    a_{1,1} +  s^{n,1}_{\msf+2} a_{2,1} +\ldots + s^{n,1}_{\vsf} a_{\vsf,1} \nonumber\\
    &= s^{n,1}_1 f_{1,1}  +  s^{n,1}_{\msf+1}    a_{1,1} +  s^{n,1}_{\msf+2} a_{2,1} +\ldots + s^{n,1}_{\vsf} a_{\vsf,1},    \label{eq:Kc 1 first column}
\end{align}      
    such that in the transmitted  linear combination of worker $n$ the coefficient of $W_{1,1}$ is $0$. Since there are totally $\vsf$ variables (i.e., $a_{1,1},\ldots,a_{\vsf,1}$) and $\vsf$ linear constraints over these variables whose
      coefficients  are uniformly i.i.d. over   $\mathbb{F}_{\qsf}$, we can solve these $\vsf$ variables with high probability. By considering all the columns in ${\bf F^{\prime}} $, we can guarantee  that in the transmitted linear combination of each worker, the coefficients of the sub-messages which it cannot compute are $0$. Moreover,  for each set $\Ac \subseteq [\Nsf]$ where $|\Ac|=\Nsf_{\rm r}$, the $\Nsf_{\rm r}$ vectors, $\sv^{\Ac(1),1},\ldots,\sv^{\Ac(\Nsf_{\rm r}),1}$, are linearly independent with high probability. Hence,   the master can recover ${\bf F^{\prime}} [W_{1,1};W_{2,1};\ldots; W_{\Ksf,1};W_{1,2};\ldots;W_{\Ksf,\msf}]$ from the answer of workers in $\Ac$.
      
It was proved in~\cite{efficientgradientcoding} that when $\Ksf_{\rm c}=1$, the   communication cost $\frac{\Nsf_{\rm r}}{\msf}$ is optimal under the constraint of linear coding in  the computing phase and symmetric transmission (i.e., the number of symbols transmitted by each worker is the same).

  \subsubsection{$\msf=1$}
\label{subsub:review m=1}
We then review the computing scheme in~\cite{linearcomput2020wan} for the case where $\msf=1$. Here we focus on the regime where $\frac{\Ksf}{\Nsf} < \Ksf_{\rm c} \leq \frac{\Ksf}{\Nsf}\Nsf_{\rm r}$, because   the remaining regimes of $\Ksf_{\rm c}$  can be solved by an  extension of the computing scheme in~\cite{linearcomput2020wan} for the above considered regime. 
 The cyclic assignment is also used for the data assignment phase.
 In the computing phase, the main idea is to let each worker send $\frac{\Ksf}{\Nsf}$ linear combinations of  messages, such that the master receives $\Nsf_{\rm r}\frac{\Ksf}{\Nsf}$ linear combinations of messages, among which it   then recovers the $\Ksf_{\rm c}$ desired ones. 
We generate $\vsf=\frac{\Ksf}{\Nsf}\Nsf_{\rm r}-\Ksf_{\rm c}$ virtually demanded linear combinations of messages, such that the effective   demand matrix   is 
\begin{align}
{\bf F^{\prime}} =\begin{bmatrix}       
f_{1,1} &\cdots &f_{1,\Ksf}\\
\vdots& \ddots & \vdots \\
f_{\Ksf_{\rm c},1} &\cdots &f_{\Ksf_{\rm c},\Ksf}\\ 
  a_{1,1}&\cdots  &   a_{1,\Ksf}   \\
\vdots &\ddots & \vdots & \\
      a_{\vsf,1}&\cdots  &   a_{\vsf,\Ksf}  
\end{bmatrix}. \label{eq:m=1 F prime}
\end{align}
 Different from the computing scheme in~\cite{efficientgradientcoding,adaptiveGC2020} for the case  $\Ksf_{\rm c}=1$ where the transmission vectors of workers  are first randomly picked, 
 the computing scheme in~\cite{linearcomput2020wan} first choose the value  of each $a_{i,k}$ where $i\in [\vsf]$ and $k\in [\Ksf]$ uniformly i.i.d over $\mathbb{F}_{\qsf}$. The next step is to determine the transmission vectors  of each worker $n\in [\Nsf]$, denoted by $\sv^{n,j}$ for $j \in \left[ \frac{\Ksf}{\Nsf} \right]$,  where the $j^{\text{th}}$ transmitted linear combination by worker $n$ is 
\begin{align}
\sv^{n,j} \ {\bf F^{\prime}}  \ [W_{1};W_{2};\ldots; W_{\Ksf}].
\end{align}
Notice that the number of messages which worker $n$ cannot compute is $|[\Ksf]\setminus \Zc_n|= \frac{\Ksf}{\Nsf}(\Nsf_{\rm r}-1)$. The sub-matrix of ${\bf F^{\prime}}$ including the   columns with the indices in $[\Ksf]\setminus \Zc_n$ has the dimension $\frac{\Ksf}{\Nsf}\Nsf_{\rm r}  \times \frac{\Ksf}{\Nsf}(\Nsf_{\rm r}-1)$. 
Since the elements in this sub-matrix are uniformly i.i.d. over $\mathbb{F}_{\qsf}$,   a vector basis for the left-side null space of this sub-matrix is the set of $\frac{\Ksf}{\Nsf}$ linearly  independent vectors with high probability. Hence, we let $\sv^{n,j}$ where $j \in \left[ \frac{\Ksf}{\Nsf} \right]$ be each of this left-side null space vector, such that in the linear combination  $\sv^{n,j} \ {\bf F^{\prime}}  \ [W_{1};W_{2};\ldots; W_{\Ksf}]$
 the coefficients  of the messages which worker $n$ cannot compute are $0$. It was also proved in~\cite{linearcomput2020wan} that for each set $\Ac \subseteq [\Nsf]$ where $|\Ac|=\Nsf_{\rm r}$, the set of vectors $\sv^{n,j}$ where $n \in \Ac$ and $j \in \left[ \frac{\Ksf}{\Nsf} \right]$ are linearly independent with high probability, such that the master can recover ${\bf F^{\prime}}  \ [W_{1};W_{2};\ldots; W_{\Ksf}]$ from the answer of workers in $\Ac$. 
 
 The   communication cost by the computing scheme in~\cite{linearcomput2020wan}  is  $\Nsf_{\rm r}\Ksf_{\rm c}$ when $\Ksf_{\rm r} \leq \frac{\Ksf}{\Nsf}$; is $\frac{\Ksf \Nsf_{\rm r}}{\Nsf}$ when $\frac{\Ksf}{\Nsf} \leq \Ksf_{\rm c}\leq \frac{\Ksf}{\Nsf} \Nsf_{\rm r}$; is $\Ksf_{\rm c}$ when $\Ksf_{\rm c} \geq \frac{\Ksf}{\Nsf}\Nsf_{\rm r}$. The   communication cost is exactly optimal when $\Ksf=\Nsf$, or when $\Ksf_{\rm c} \in \left[ \left\lceil \frac{\Ksf}{\binom{\Nsf}{\Nsf-\Nsf_{\rm r}+1}} \right\rceil  \right]$, or when  $\Ksf_{\rm c} \in \left[  \frac{\Ksf}{\Nsf}\Nsf_{\rm r}  : \Ksf \right]$. In addition, it is   optimal under the constraint of cyclic assignment when $\Nsf$ divides $\Ksf$.

\section{Main Results}
\label{sec:main}
In this section, we present our novel results in this paper. We first provide a converse bound under the constraint of cyclic assignment, which will be proved in Section~\ref{sec:converse}.
\begin{thm}
\label{thm:converse}
 For the  $(\Ksf,\Nsf,\Nsf_{\rm r}, \Ksf_{\rm c}, \msf)$ distributed linearly separable computation problem,
\begin{itemize}
\item when $\Ksf_{\rm c} \in \left[ \frac{\Ksf}{\Nsf} (\Nsf_{\rm r}-\msf+1) \right]$, by defining $\usf:= \left\lceil \frac{\Ksf_{\rm c} \Nsf}{\Ksf}\right\rceil$, we have 
\begin{subequations}
\begin{align}
\Rsf^{\star}_{\text{cyc}} \geq   \frac{ \Nsf_{\rm r} \Ksf_{\rm c}}{\msf+\usf-1}. \label{eq:case 1 converse}
\end{align} 
\item when $\Ksf_{\rm c} \in  \left[ \frac{\Ksf}{\Nsf} (\Nsf_{\rm r}-\msf+1) :\Ksf \right] $, we have 
\begin{align}
\Rsf^{\star}_{\text{cyc}} \geq \Rsf^{\star}  \geq  \Ksf_{\rm c}. \label{eq:case 2 converse}
\end{align} 
\end{subequations}
\end{itemize}
\hfill $\square$ 
\end{thm}

We then introduce the   computation-communication costs tradeoff by the novel computing scheme in the following theorem.
\begin{thm}
\label{thm:achie}
 For the  $(\Ksf,\Nsf,\Nsf_{\rm r}, \Ksf_{\rm c}, \msf)$ distributed linearly separable computation problem where  
\begin{align}
40 \geq \Nsf \geq \frac{\msf+\usf-1}{\usf}+ \usf(\Nsf_{\rm r}-\msf-\usf+1), \label{eq:key constraint}
\end{align}  
 the computation-communication costs tradeoff   $(\msf,\Rsf_{{\rm ach}})$ is achievable, where 
\begin{itemize}
 \item when  $\Ksf_{\rm c} \in \left[\frac{\Ksf}{\Nsf} \right]$, 
 \begin{subequations}
\begin{align}
  \Rsf_{{\rm ach}}= \frac{\Ksf_{\rm c} \Nsf_{\rm r}}{\msf} \label{eq:achie case 1}
\end{align} 
\item when  $ \Ksf_{\rm c}  \in \left[ \frac{\Ksf}{\Nsf}: \frac{\Ksf}{\Nsf}(\Nsf_{\rm r}-\msf+1) \right]  $,
\begin{align}
\Rsf_{{\rm ach}}=\frac{\Nsf_{\rm r} \Ksf \usf}{\Nsf(\msf+\usf-1)}  ; \label{eq:achie case 2}
\end{align}
\item when  $ \Ksf_{\rm c} \in \left[ \frac{\Ksf}{\Nsf}(\Nsf_{\rm r}-\msf+1) : \Ksf \right]$,
\begin{align}
\Rsf_{{\rm ach}}=  \Ksf_{\rm c}. \label{eq:achie case 3}
\end{align}
\label{eq:achieve three cases}
\end{subequations}
\end{itemize}
\hfill $\square$ 
\end{thm}
 Notice that the RHS of   the  constraint~\eqref{eq:key constraint} 
\begin{align}
\Nsf \geq \frac{\msf+\usf-1}{\usf}+ \usf(\Nsf_{\rm r}-\msf-\usf+1), \label{eq:key part constraint}
\end{align} 
  will be explained   in Remark~\ref{rem:intuition of achie constraint} from a viewpoint of linear space dimension.  
It can be seen that in the first case of the proposed computing scheme (i.e., $\Ksf_{\rm c} \in \left[\frac{\Ksf}{\Nsf} \right]$), we have $\usf=1$ and thus the constraint~\eqref{eq:key part constraint}  always holds.   
In the third case of the proposed computing scheme (i.e., $\Ksf_{\rm c} \in \left[ \frac{\Ksf}{\Nsf}(\Nsf_{\rm r}-\msf+1) : \Ksf \right]$), 
  we have $\usf \geq \Nsf_{\rm r}-\msf+1$ and thus the constraint in~\eqref{eq:key part constraint} always holds.

 While proving the decodability of the proposed computing scheme in Theorem~\ref{thm:achie}, we use  the Schwartz-Zippel lemma~\cite{Schwartz,Zippel,Demillo_Lipton} in Appendix~\ref{sec:proof of encoding and decoding lemma}. For  the non-zero polynomial condition for the   Schwartz-Zippel lemma, we   numerically verify   all cases that $40 \geq \Nsf \geq \frac{\msf+\usf-1}{\usf}+ \usf(\Nsf_{\rm r}-\msf-\usf+1)$,  and conjecture in the rest of the paper that the condition holds for any case where $\Nsf \geq \frac{\msf+\usf-1}{\usf}+ \usf(\Nsf_{\rm r}-\msf-\usf+1)$, i.e., in Theorem~\ref{thm:achie} we replace the constraint~\eqref{eq:key constraint} by~\eqref{eq:key part constraint}.

In Section~\ref{sec:achie}, for the sake of space limitation, we will only provide our novel computing scheme for the second case~\eqref{eq:achie case 2}   (i.e.,  $ \Ksf_{\rm c}  \in \left[ \frac{\Ksf}{\Nsf}: \frac{\Ksf}{\Nsf}(\Nsf_{\rm r}-\msf+1) \right] $). By the exactly same method as described in~\cite[Sections IV-B and IV-C]{linearcomput2020wan}, the computing schemes for the first and third cases can be   obtained by the direct extensions of the computing scheme for the second case. More precisely, 
\begin{itemize}
\item $\Ksf_{\rm c} \in \left[\frac{\Ksf}{\Nsf} \right]$. When $\Ksf_{\rm c}=1$, it can be easily shown (see~\cite[Section IV-B]{linearcomput2020wan}) that the $ (\Ksf,\Nsf,\Nsf_{\rm r}, 1 , \msf)$ distributed linearly separable computation problem is equivalent to   
the $ (\Nsf,\Nsf,\Nsf_{\rm r}, 1 , \msf)$ distributed linearly separable computation problem, which needs the communication cost $\frac{  \Nsf_{\rm r}}{\msf}$ from~\eqref{eq:achie case 2}. For  $\Ksf_{\rm c} \in \left[2:\frac{\Ksf}{\Nsf} \right]$, we can treat the $(\Ksf,\Nsf,\Nsf_{\rm r}, \Ksf_{\rm c}, \msf)$ distributed linearly separable computation problem as $\Ksf_{\rm c}$ independent  $ (\Ksf,\Nsf,\Nsf_{\rm r}, 1 , \msf)$ distributed linearly separable computation problems; thus the   communication cost is $\frac{ \Ksf_{\rm c} \Nsf_{\rm r}}{\msf}$, coinciding with~\eqref{eq:achie case 1}. 
\item $ \Ksf_{\rm c} \in \left[ \frac{\Ksf}{\Nsf}(\Nsf_{\rm r}-\msf+1) : \Ksf \right]$. When  $ \Ksf_{\rm c}=\frac{\Ksf}{\Nsf}(\Nsf_{\rm r}-\msf+1) $,  from~\eqref{eq:achie case 2} it can be seen that the   communication cost is $\frac{\Nsf_{\rm r} \Ksf \usf}{\Nsf(\msf+\usf-1)} = \frac{\Ksf \usf}{\Nsf}=\Ksf_{\rm c}$, coinciding with~\eqref{eq:achie case 3}. When $ \Ksf_{\rm c}>\frac{\Ksf}{\Nsf}(\Nsf_{\rm r}-\msf+1) $, as in~\cite[Section IV-C]{linearcomput2020wan}, we can divide each demanded linear combination into $\binom{\Ksf_{\rm c}-1}{\frac{\Ksf}{\Nsf}(\Nsf_{\rm r}-\msf+1)-1}$ equal-length  sub-combinations, each of which has $\frac{\Lsf}{\binom{\Ksf_{\rm c}-1}{\frac{\Ksf}{\Nsf}(\Nsf_{\rm r}-\msf+1)-1}} $ symbols.
 We then treat the $(\Ksf,\Nsf,\Nsf_{\rm r}, \Ksf_{\rm c}, \msf)$ distributed linearly separable computation problem as $\binom{\Ksf_{\rm c}}{\frac{\Ksf}{\Nsf}(\Nsf_{\rm r}-\msf+1)}$ independent  $\big(\Ksf,\Nsf,\Nsf_{\rm r}, \frac{\Ksf}{\Nsf}(\Nsf_{\rm r}-\msf+1), \msf \big)$ distributed linearly separable computation sub-problems, where in each sub-problem we let the master recover $\frac{\Ksf}{\Nsf}(\Nsf_{\rm r}-\msf+1)$ sub-combinations, with the communication cost $ \frac{\frac{\Ksf}{\Nsf}(\Nsf_{\rm r}-\msf+1) }{\binom{\Ksf_{\rm c}-1}{\frac{\Ksf}{\Nsf}(\Nsf_{\rm r}-\msf+1)-1} }$; thus the total communication cost is $$
\binom{\Ksf_{\rm c}}{\frac{\Ksf}{\Nsf}(\Nsf_{\rm r}-\msf+1)}  \frac{\frac{\Ksf}{\Nsf}(\Nsf_{\rm r}-\msf+1) }{\binom{\Ksf_{\rm c}-1}{\frac{\Ksf}{\Nsf}(\Nsf_{\rm r}-\msf+1)-1} } = \Ksf_{\rm c},
$$
coinciding with~\eqref{eq:achie case 3}. 
\end{itemize}

By comparing the proposed converse bound in Theorem~\ref{thm:converse} and the proposed scheme in Theorem~\ref{thm:achie}, we can directly have the following (order) optimality results.
\begin{thm}
\label{thm:optimality}
For the  $(\Ksf,\Nsf,\Nsf_{\rm r}, \Ksf_{\rm c}, \msf)$ distributed linearly separable computation problem where $\Nsf \geq \frac{\msf+\usf-1}{\usf}+ \usf(\Nsf_{\rm r}-\msf-\usf+1)$,
    \begin{itemize}
 \item when $\Ksf=\Nsf$, we have 
    \begin{align}
  \Rsf^{\star}_{\text{cyc}}=\Rsf_{{\rm ach}}= \begin{cases} \frac{\Nsf_{\rm r}\Ksf_{\rm c} }{\msf+\Ksf_{\rm c}-1}, & \text{ if }  \Ksf_{\rm c} \in [\Nsf_{\rm r}-\msf+1]; \\ \Ksf_{\rm c},  & \text{ if }  \Ksf_{\rm c} \in [\Nsf_{\rm r}-\msf+1: \Ksf] ; \end{cases} \label{eq:opt N=K}
    \end{align}
\item when $\Ksf_{\rm c} \in \left[\frac{\Ksf}{\Nsf} \right]$, we have 
\begin{align}
\Rsf^{\star}_{\text{cyc}}=\Rsf_{{\rm ach}}=\frac{\Nsf_{\rm r}\Ksf_{\rm c} }{\msf}; \label{eq:opt reg1}
\end{align} 
\item  when $  \Ksf_{\rm c} \in \left[ \frac{\Ksf}{\Nsf}+1: \frac{\Ksf}{\Nsf}(\Nsf_{\rm r}-\msf+1)-1 \right]$, we have 
\begin{align}
\Rsf^{\star}_{\text{cyc}} \geq  \frac{\Ksf_{\rm c}}{\frac{\Ksf}{\Nsf}\usf}  \Rsf_{{\rm ach}} \geq \frac{\Rsf_{{\rm ach}} }{2}; \label{eq:opt reg2}
\end{align} 
  \item when $\Ksf_{\rm c} \in  \left[\frac{\Ksf}{\Nsf}(\Nsf_{\rm r}-\msf+1 ): \Ksf \right]$, we have 
   \begin{align}
    \Rsf^{\star}=\Rsf^{\star}_{\text{cyc}}=\Rsf_{{\rm ach}}=\Ksf_{\rm c}; \label{eq:opt reg3}
      \end{align}
  \end{itemize}
  \hfill $\square$ 
\end{thm}
In words, for the considered problem satisfying the constraint in~\eqref{eq:key part constraint}, when $\Ksf_{\rm c} \in  [\Nsf_{\rm r}-\msf+1: \Ksf]$, the proposed computing scheme is exactly optimal;
when $\Ksf=\Nsf$ or $\Ksf_{\rm c} \in \left[\frac{\Ksf}{\Nsf} \right]$, the proposed computing scheme is optimal under the constraint of cyclic assignment; when $\Nsf$ divides $\Ksf$ and  $\Ksf_{\rm c} \in \left[ \frac{\Ksf}{\Nsf}+1: \frac{\Ksf}{\Nsf}(\Nsf_{\rm r}-\msf+1)-1 \right]$, the proposed scheme is order optimal within a factor of $\frac{\frac{\Ksf}{\Nsf}\usf}{\Ksf_{\rm c}}\leq  2$ under the constraint of cyclic assignment. 
 
Notice that when $\Ksf_{\rm c} =1$, the proposed computing scheme achieves the same communication load as in~\cite{efficientgradientcoding,adaptiveGC2020}, which was proved to be optimal under the constraint of   linear coding in  the computing phase and symmetric transmission. Instead, in this paper we prove  that it is optimal only under the constraint of cyclic assignment.

\begin{figure}
    \centering
    \begin{subfigure}[t]{0.5\textwidth}
        \centering
        \includegraphics[scale=0.6]{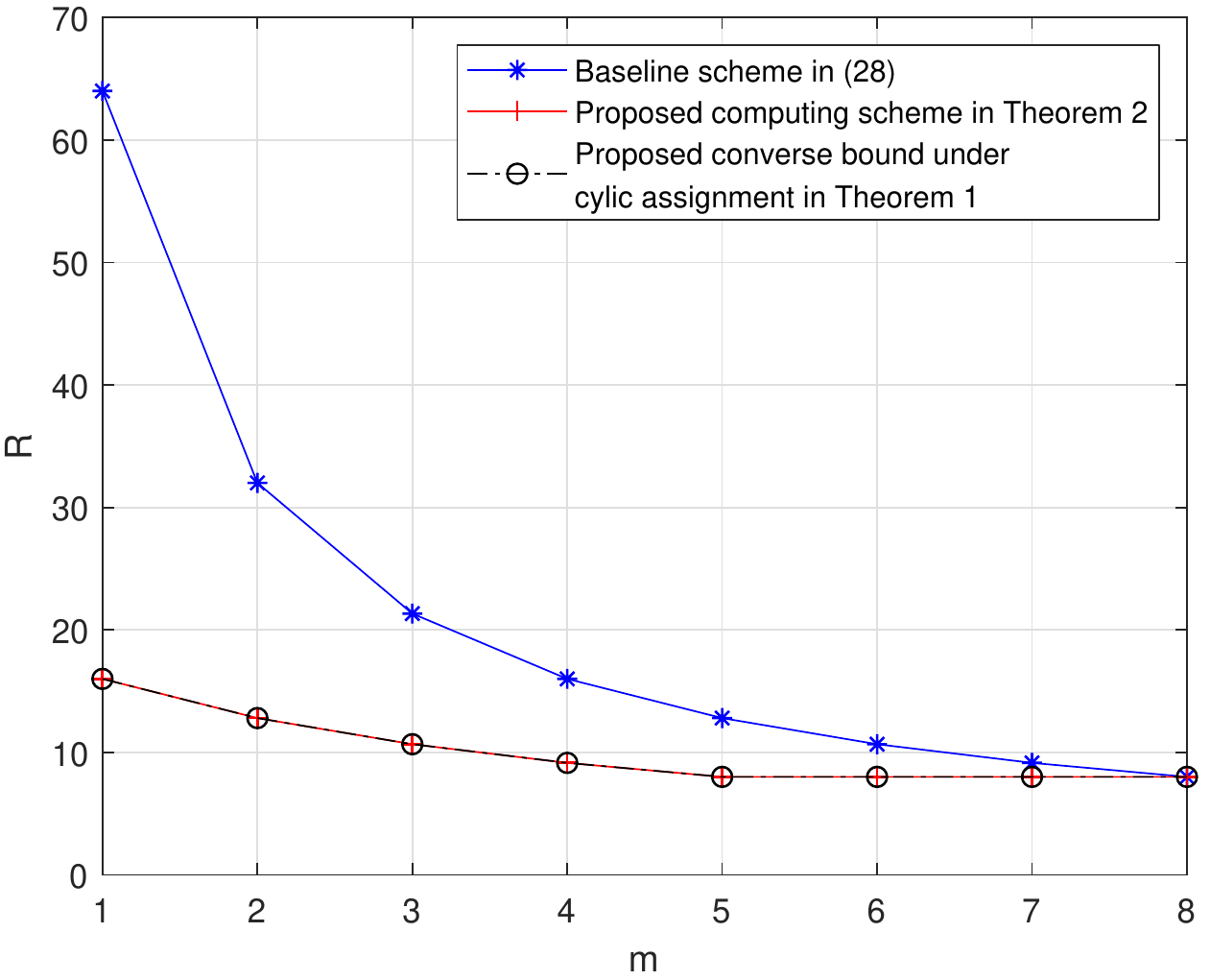}
        \caption{\small The computation-communication costs tradeoff for the case $\Ksf=20$, $\Nsf=10$, $\Nsf_{\rm r}=8$, $\Ksf_{\rm c}=8$.}
        \label{fig:numerical 1a}
    \end{subfigure}%
    ~ 
    \begin{subfigure}[t]{0.5\textwidth}
        \centering
        \includegraphics[scale=0.6]{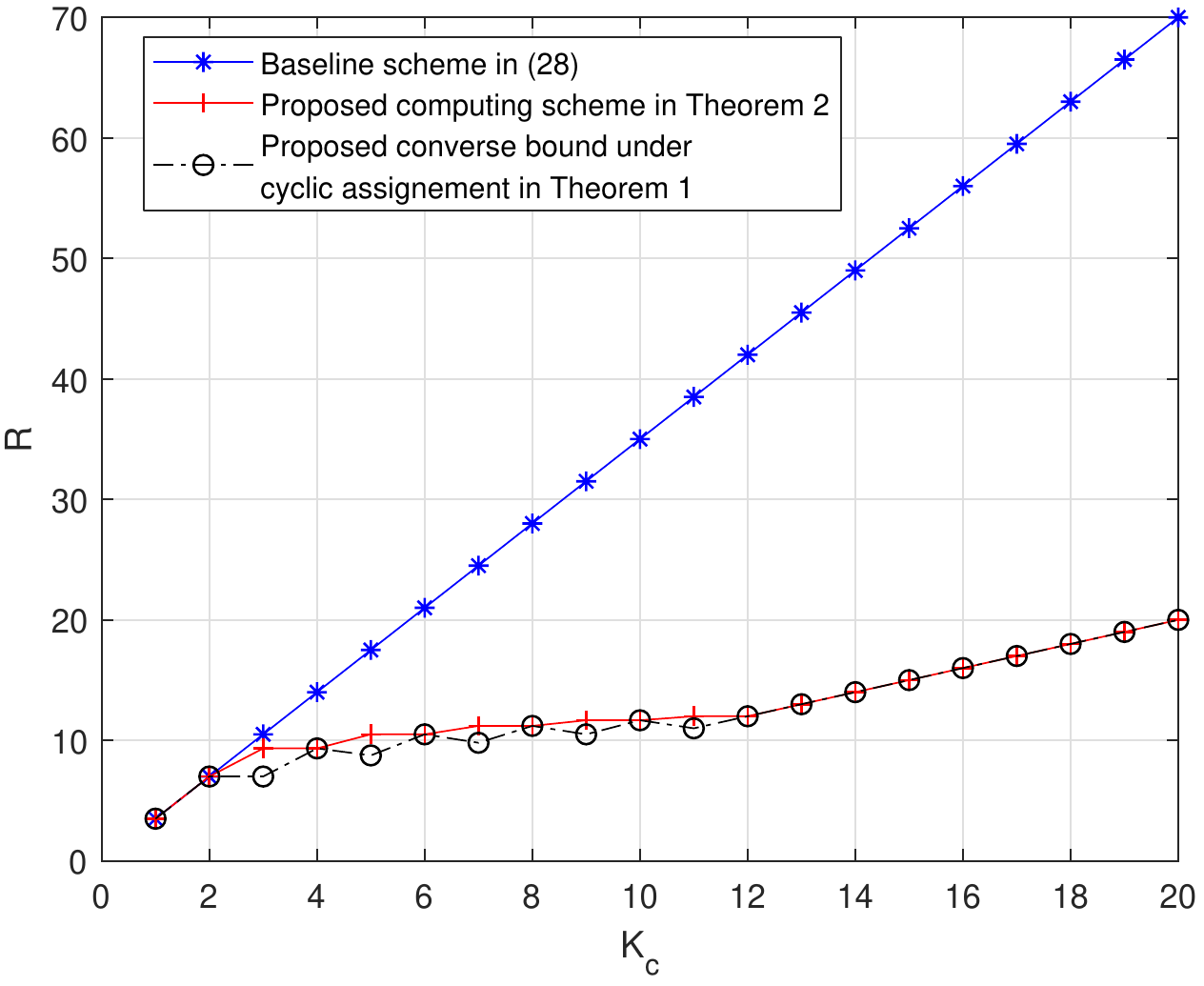}
        \caption{\small The    communication costs  for the case $\Ksf=20$, $\Nsf=10$, $\Nsf_{\rm r}=7$, $\msf=2$, $\Ksf_{\rm c} \in [20]$.}
        \label{fig:numerical 1b}
    \end{subfigure}
    \caption{\small Numerical evaluations for the considered distributed linearly separable computation problem.}
    \label{fig:numerical 1}
\end{figure}

 In Fig.~\ref{fig:numerical 1}, we   provide some numerical evaluations on the proposed converse and achievable bounds. For the sake of comparison, we introduce a baseline scheme. For the case where $\Ksf_{\rm c}=1,$ 
   the computing scheme in~\cite{efficientgradientcoding,adaptiveGC2020} (reviewed in Section~\ref{sub:review of two cases}) needs the communication cost $\frac{\Nsf_{\rm r}}{\msf}$ for each $\msf \in [\Nsf_{\rm}]$. Hence, a simple baseline scheme can be obtained by treating the considered problem as $\Ksf_{\rm c}$ independent sub-problems, where in each sub-problem the master recover one of its desired linear combination. Thus the   communication cost     for the baseline scheme is 
   \begin{align}
  \Rsf_{\text{base}} =  \frac{\Ksf_{\rm c}\Nsf_{\rm r}}{\msf}, \ \forall \msf \in [\Nsf_{\rm r}].\label{eq:baseline scheme}
   \end{align}
   
In Fig.~\ref{fig:numerical 1a}, we consider the  distributed linearly separable computation problem where  $\Ksf=20$, $\Nsf=10$, $\Nsf_{\rm r}=8$, and $\Ksf_{\rm c}=8$. In this example, the constraint in~\eqref{eq:key part constraint} always holds. It can be seen from Fig.~\ref{fig:numerical 1a} that the proposed computing scheme outperforms the baseline scheme and coincides with the proposed converse bound. 

In Fig.~\ref{fig:numerical 1b},  we consider the  distributed linearly separable computation problem where  $\Ksf=20$, $\Nsf=10$, $\Nsf_{\rm r}=7$, $\msf=2$. For each $\Ksf_{\rm c} \in [20]$, we plot the   communication costs. In this example, 
the constraint in~\eqref{eq:key part constraint} also always holds. It can be seen from Fig.~\ref{fig:numerical 1b} that the proposed computing scheme outperforms the baseline scheme. The propose scheme    coincides with the proposed converse bound when  $\Ksf_{\rm c} \leq \frac{\Ksf}{\Nsf}=2$, or when  $\Ksf_{\rm c}$ divides $\frac{\Ksf}{\Nsf}$, or when $\Ksf_{\rm c} \geq \frac{\Ksf}{\Nsf}(\Nsf_{\rm r}-\msf+1)=12$.

\section{Proof of Theorem~\ref{thm:converse}}
\label{sec:converse}
As shown in~\cite[Section II]{linearcomput2020wan}, since the elements of the demand matrix ${\bf F}$ are uniformly i.i.d. over larger enough field $\mathbb{F}_{\qsf}$, a simple cut-set bound argument yields
 \begin{align}
\Rsf^{\star}_{\text{cyc}} \geq \Rsf^{\star}  \geq  \Ksf_{\rm c},\label{eq:trivial lower bound}
 \end{align}
 which coincides with the converse bound in~\eqref{eq:case 2 converse} for the case $\Ksf_{\rm c} \in  \left[ \frac{\Ksf}{\Nsf} (\Nsf_{\rm r}-\msf+1) :\Ksf \right] $. 
Hence, in the following we focus on the case  $\Ksf_{\rm c} \in \left[ \frac{\Ksf}{\Nsf} (\Nsf_{\rm r}-\msf+1) \right]$.

We  will use an example to illustrate the main idea. 
\begin{example}
\label{ex:converse example}
In this example, we have $\Nsf=\Ksf=5$, $\Nsf_{\rm r}=4$, $\msf=2$, and $\Ksf_{\rm c}=2$. Hence, the number of datasets assigned to each worker is $\Msf=\frac{\Ksf}{\Nsf}(\Nsf-\Nsf_{\rm r}+\msf)=3$. Each dataset is assigned to $3$ workers. 
With the cyclic assignment,  we assign
\begin{align*}
\begin{array}{rl|c|cc|c|cc|c|cc|c|cc|c|}\cline{3-3}\cline{6-6}\cline{9-9}\cline{12-12}\cline{15-15}
&&\rule{0pt}{1.2em}\mbox{Worker 1} &&&\rule{0pt}{1.2em}\mbox{Worker 2} &&& \rule{0pt}{1.2em}\mbox{Worker 3} &&& \rule{0pt}{1.2em}\mbox{Worker 4} &&& \rule{0pt}{1.2em}\mbox{Worker 5}  \\\cline{3-3}\cline{6-6}\cline{9-9}\cline{12-12}\cline{15-15}
&& D_1& &\mbox{\tiny }& D_2 && & D_3&& & D_4 && & D_5\\
&& D_2& &\mbox{\tiny }& D_3 && & D_4&& & D_5 && & D_1\\
&& D_3& &\mbox{\tiny }& D_4 && & D_5&& & D_1 && & D_2\\
\cline{3-3}\cline{6-6}\cline{9-9}\cline{12-12}\cline{15-15}
\end{array}
\end{align*}
We consider the demand matrix ${\bf F}$ whose dimension is $2\times 5$ with  elements uniformly i.i.d. over large field $\mathbb{F}_{\qsf}$. Hence, the sub-matrix including each $\Kc_{\rm c}=2$ columns is full-rank with high probability.

Notice that in this example the number of stragglers is $\Nsf-\Nsf_{\rm r} =1 $.
We first consider    that  worker $5$ is the straggler; thus the master should recover ${\bf F} [W_1;\ldots;W_5]$ from the answers of workers in $\Ac=[4]$. 
 In addition, each dataset is assigned to $\Nsf-\Nsf_{\rm r}+\msf=3$ workers. Hence, there must exist one dataset assigned to all the straggler(s) which is also assigned to   $\msf$ responding workers.  In this example, all of $D_1$, $D_2$, and $D_5$   belong to such datasets. Now we select one of them, e.g., $D_2$. Note that $D_2$ is assigned to workers $\Hc_2=\{1,2,5\}$. 
We then consider the next dataset $D_{\text{Mod}(2+1,\Ksf)}=D_3$. The workers storing dataset $D_3$ (denoted by $\Hc_3$) is obtained by right-shifting  $\Hc_2$ by one position, i.e., $\Hc_3=\{1,2,3\}$. Hence, there is exactly one new worker in $\Hc_3$ who is not in $\Hc_2 \cap \Ac$, which is worker $3$. 
So we have  
$$
|(\Hc_{2}\cup \Hc_{3}) \cap \Ac|=\msf+(2-1)=3=\msf+\Ksf_{\rm c}-1;
$$
in other words, in the set of responding workers $\Ac$, the number of workers who can compute $W_2$ or $W_3$ is equal to $3$.
In addition, the sub-matrix of ${\bf F}$ including the columns in $\{2,3\}$ is full-rank (with rank $\Ksf_{\rm c}=2$). Recall that    each message has $\Lsf$ uniformly i.i.d. symbols. Hence, the number of transmitted symbols by workers in  $(\Hc_{2}\cup \Hc_{3}) \cap \Ac$ should be no less than $2 \Lsf$; thus 
\begin{align}
\sum_{n\in \big((\Hc_{2}\cup \Hc_{3}) \cap \Ac \big) } T_{n} =T_{1}+T_{2}+T_{3} \geq \Ksf_{\rm c} \Lsf=2 \Lsf. \label{eq:134}
\end{align}

Similarly, considering that  worker $4 $ is the straggler,  we have 
\begin{align}
 T_{5}+T_{1}+T_{2} \geq  \Ksf_{\rm c} \Lsf=2 \Lsf. \label{eq:235}
\end{align}
Considering that  worker $3$ is the straggler,  we have 
\begin{align}
 T_{4}+T_{5}+T_{1} \geq  \Ksf_{\rm c} \Lsf=2 \Lsf. \label{eq:124}
\end{align}
Considering that  worker $2$ is the straggler,  we have 
\begin{align}
 T_{3}+T_{4}+T_{5} \geq  \Ksf_{\rm c} \Lsf=2 \Lsf. \label{eq:513}
\end{align}
Considering that  worker $1$ is the straggler,  we have 
\begin{align}
 T_{2}+T_{3}+T_{4} \geq  \Ksf_{\rm c} \Lsf=2 \Lsf. \label{eq:452}
\end{align}

By summing~\eqref{eq:134}-\eqref{eq:452}, we have 
\begin{align}
 T_1+T_2+T_3+T_4+T_5\geq \frac{10}{3} \Lsf,
\end{align}
 which leads that 
 \begin{align}
 \Rsf^{\star}_{\text{cyc}} \geq \max_{\Ac \subseteq [5]:|\Ac|=\Nsf_{\rm r}=4} \frac{ \sum_{j\in \Ac} T_{j}}{\Lsf}
  \geq \frac{8}{3},
 \end{align}
  as the converse bound in~\eqref{eq:case 1 converse}.
   \hfill $\square$ 
\end{example}

We are now ready to generalize the proposed converse bound under the constraint of cyclic assignment in Example~\ref{ex:converse example}. Recall that we consider the case where $\Ksf_{\rm c} \in \left[ \frac{\Ksf}{\Nsf} (\Nsf_{\rm r}-\msf+1) \right]$ and that  $\usf= \left\lceil \frac{\Ksf_{\rm c} \Nsf}{\Ksf}\right\rceil$. 
The demand matrix ${\bf F}$ has dimension  $\Ksf_{\rm c} \times \Ksf$ with  elements uniformly i.i.d. over large field. Hence, the sub-matrix including each $\Kc_{\rm c}$ columns is full-rank with high probability.
By the cyclic assignment, as shown in~\eqref{eq:Hk in cyclic assignment}, each dataset $D_k$ is assigned to workers 
$\Hc_k= \big\{\text{Mod}(k,\Nsf),  \text{Mod}(k-1,\Nsf),\ldots,  \text{Mod}(k-\Nsf+\Nsf_{\rm r}-\msf+1, \Nsf ) \big\}.$

We  consider the set of stragglers whose are adjacent. Thus each time we choose one integer $n \in [\Nsf]$, let $\Sc_n:= \{\text{Mod}(n,\Nsf), \text{Mod}(n-1,\Nsf),\ldots, \text{Mod}(n-\Nsf+\Nsf_{\rm r}+1,\Nsf)\}$ where $|\Sc_n|=\Nsf-\Nsf_{\rm r}$, be the set of stragglers.
  The master should recover ${\bf F} [W_1;\ldots;W_{\Ksf}]$ from the answers of workers in $[\Nsf] \setminus \Sc_n$. 
  From the cyclic assignment, there are exactly $\frac{\Ksf}{\Nsf}$ datasets, denoted by $\Uc_0=\left\{\text{Mod}(n+\msf,\Nsf)+ p \Nsf  : p \in \left[0:  \frac{\Ksf}{\Nsf} -1 \right] \right\}$, which are exclusively  assigned to the workers in 
\begin{align*}
& \Hc_{\Uc_0} = \Sc_n \cup \{\text{Mod}(n+1,\Nsf),\text{Mod}(n+2,\Nsf),\ldots,\text{Mod}(n+\msf,\Nsf)\}\\
&= \{\text{Mod}(n-\Nsf+\Nsf_{\rm r}+1,\Nsf), \text{Mod}(n-\Nsf+\Nsf_{\rm r}+2,\Nsf) ,\ldots,\text{Mod}(n+\msf,\Nsf) \};
\end{align*}    
    Then for each $i \in [\usf-1]$, the   datasets in 
$
\Uc_i  = \left\{\text{Mod}(n+\msf+i,\Nsf)+ p \Nsf  : p \in \left[0:  \frac{\Ksf}{\Nsf} -1 \right] \right\}, 
$    
         are exclusively  assigned to the workers in  
\begin{align*}
      \Hc_{\Uc_i}= \{\text{Mod}(n-\Nsf+\Nsf_{\rm r}+i+1,\Nsf), \text{Mod}(n-\Nsf+\Nsf_{\rm r}+i+2,\Nsf) ,\ldots,\text{Mod}(n+\msf+i,\Nsf) \}.
\end{align*}  
It can be seen that there are totally $\frac{\Ksf}{\Nsf} \usf $  datasets in
 $\cup_{i\in [0: \usf-1]} \Uc_i $, which are exclusively  assigned to the workers in
 \begin{align*}
& \cup_{i\in [0: \usf-1]}       \Hc_{\Uc_i}=  \{\text{Mod}(n-\Nsf+\Nsf_{\rm r}+1,\Nsf), \text{Mod}(n-\Nsf+\Nsf_{\rm r}+2,\Nsf) ,\ldots,\text{Mod}(n+\msf+\usf-1,\Nsf) \}   \\
&= \Sc_n \cup \{ \text{Mod}(n+1,\Nsf) ,\ldots,\text{Mod}(n+\msf+\usf-1,\Nsf)\}.
 \end{align*}
  Note that since $\usf \leq \Nsf_{\rm r}-\msf+1$, we have $ \Sc_n \cap \{ \text{Mod}(n+1,\Nsf) ,\ldots,\text{Mod}(n+\msf+\usf-1,\Nsf)\} =\emptyset$.
  In other words, the number of responding workers in $\cup_{i\in [0: \usf-1]}       \Hc_{\Uc_i}$ is  
  $$
  \left|\left(\cup_{i\in [0: \usf-1]}       \Hc_{\Uc_i} \right)\cap ([\Nsf] \setminus \Sc_n)\right|=|\{ \text{Mod}(n+1,\Nsf) ,\ldots,\text{Mod}(n+\msf+\usf-1,\Nsf)\}|=\msf+\usf-1.
  $$
   Since $\frac{\Ksf}{\Nsf} \usf \geq \Ksf_{\rm c}$,  the sub-matrix of the demand matrix including the columns in $\cup_{i\in [0: \usf-1]} \Uc_i$  has a rank equal to $\Ksf_{\rm c}$. Hence, the number of transmitted symbols by workers in  $\{ \text{Mod}(n+1,\Nsf) ,\ldots,\text{Mod}(n+\msf+\usf-1,\Nsf)\}$ should be no less than $\Ksf_{\rm c} \Lsf$; thus 
  \begin{align}
\sum_{j\in \{ \text{Mod}(n+1,\Nsf) ,\ldots,\text{Mod}(n+\msf+\usf-1,\Nsf)\} } T_{j}  \geq \Ksf_{\rm c} \Lsf. \label{eq:general n}
\end{align}

By considering all $n \in [\Nsf]$ and summing all the inequalities as in~\eqref{eq:general n}, we have
  \begin{align}
\sum_{j\in[\Nsf] } T_{j}  \geq \frac{ \Nsf \Ksf_{\rm c}}{\msf+\usf-1} \Lsf, \label{eq:all n}
\end{align}
which leads that 
\begin{align}
\Rsf^{\star}_{\text{cyc}} \geq    \frac{\max_{\Ac \subseteq [\Nsf]:|\Ac|=\Nsf_{\rm r}} \sum_{j\in \Ac} T_{j}}{\Lsf}\geq  \frac{ \Nsf_{\rm r} \Ksf_{\rm c}}{\msf+\usf-1}, \label{eq: all nr}
\end{align}
as the converse bound  in~\eqref{eq:case 1 converse}.

\section{Proof of~\eqref{eq:achie case 2}}
\label{sec:achie}
We focus on the case where  $ \Ksf_{\rm c}  \in \left[ \frac{\Ksf}{\Nsf}: \frac{\Ksf}{\Nsf}(\Nsf_{\rm r}-\msf+1) \right]$. 
We first illustrate the main idea in the following example.
\begin{example}
\label{ex:achie example n=k first reg}

In this example, we have $\Nsf=\Ksf=6$, $\Nsf_{\rm r}=5$, $\msf=2$, and $\Ksf_{\rm c}=2$.  Since $\Nsf=\Ksf$ in this example, we have $\usf=\Ksf_{\rm c}=2$. We assume the demand matrix is 
\begin{align}
{\bf F}=
\begin{bmatrix}
f_{1,1}& f_{1,2} &f_{1,3} &f_{1,4} &f_{1,5} &f_{1,6} \\
f_{2,1}& f_{2,2} &f_{2,3} &f_{2,4} &f_{2,5} &f_{2,6} 
\end{bmatrix}
= 
\begin{bmatrix}
1 & 1 & 1 & 1 &1 &1 \\
0 & 1 & 2 & 3 &4 &5 
\end{bmatrix}
.  
\end{align}

\paragraph*{Data assignment phase}
The number of datasets assigned to each worker is $\Msf=\frac{\Ksf}{\Nsf}(\Nsf-\Nsf_{\rm r}+\msf)=3$.
We use the cyclic assignment, to assign
\begin{align*}
\begin{array}{rl|c|cc|c|cc|c|cc|c|cc|c|cc|c|}\cline{3-3}\cline{6-6}\cline{9-9}\cline{12-12}\cline{15-15}\cline{18-18}
&&\rule{0pt}{1.2em}\mbox{Worker 1} &&&\rule{0pt}{1.2em}\mbox{Worker 2} &&& \rule{0pt}{1.2em}\mbox{Worker 3} &&& \rule{0pt}{1.2em}\mbox{Worker 4} &&& \rule{0pt}{1.2em}\mbox{Worker 5} &&& \rule{0pt}{1.2em}\mbox{Worker 6}  \\\cline{3-3}\cline{6-6}\cline{9-9}\cline{12-12}\cline{15-15}\cline{18-18}
&& D_1& &\mbox{\tiny }& D_2 && & D_3&& & D_4 && & D_5&& & D_6\\
&& D_2& &\mbox{\tiny }& D_3 && & D_4&& & D_5 && & D_6&& & D_1\\
&& D_3& &\mbox{\tiny }& D_4 && & D_5&& & D_6 && & D_1&& & D_2\\
\cline{3-3}\cline{6-6}\cline{9-9}\cline{12-12}\cline{15-15}\cline{18-18}
\end{array}
\end{align*}

\paragraph*{Computing phase}
Since the   communication cost is no less than $\Nsf_{\rm r} \frac{\Ksf_{\rm c}}{\msf+\Ksf_{\rm c}-1}=\frac{10}{3}$ from the converse bound~\eqref{eq:case 1 converse}, 
we divide each message $W_k$ where $k\in [6]$ into $\msf+\Ksf_{\rm c}-1=3$ non-overlapping and equal-length sub-messages,
$W_k=\{W_{k,j}:j\in [3]\}.$ 
 Each worker should send $\Ksf_{\rm c}=2$  linear combinations of sub-messages. 
From the answers of $\Nsf_{\rm r}=5$  workers, the master totally receives $ \Nsf_{\rm r} \Ksf_{\rm c}=10$ linear combinations of sub-messages, which contain  the desired $(\msf+\Ksf_{\rm c}-1) \Ksf_{\rm c}=6$ linear combinations. Hence, 
we generate $\vsf=10-6=4$ virtually demanded linear combinations of sub-messages; thus   the effective 
demand matrix (i.e., containing original and virtual demands) is 
\begin{align}
{\bf F^{\prime}} [W_{1,1}; \ldots ; W_{6,1} ;W_{1,2}; \ldots; W_{6,3}]
\end{align}
where ${\bf F^{\prime}}  $ has dimension $ \Nsf_{\rm r} \Ksf_{\rm c} \times \Ksf(\msf+\Ksf_{\rm c}-1)= 10 \times 18$, with the form
\begin{equation}\setstretch{1.25}
{\bf F^{\prime}} =\begin{bmatrix}  \     
\tikzmark{left1} \textcolor{white}{00}1 &\cdots & \textcolor{white}{00} 1& \tikzmark{left2} \textcolor{white}{00} 0&\cdots  & \textcolor{white}{00} 0& \tikzmark{left3} \textcolor{white}{00} 0& \cdots &\textcolor{white}{00} 0 \\
\textcolor{white}{00}0 &\cdots & \textcolor{white}{00}5& \textcolor{white}{00}0&\cdots  & \textcolor{white}{00}0& \textcolor{white}{00}0&\cdots &\textcolor{white}{00}0\\
 \textcolor{white}{00}0&\cdots  & \textcolor{white}{00}0 & \textcolor{white}{00}1 &\cdots & \textcolor{white}{00}1& \textcolor{white}{00}0&\cdots  & \textcolor{white}{00}0 \\
 \textcolor{white}{00} 0&\cdots  &\textcolor{white}{00} 0 & \textcolor{white}{00}0 &\cdots & \textcolor{white}{00}5& \textcolor{white}{00}0&\cdots  & \textcolor{white}{00}0 \\
 \textcolor{white}{00} 0&\cdots  & \textcolor{white}{00}0 &\textcolor{white}{00} 0 &\cdots & \textcolor{white}{00}0&\textcolor{white}{00} 1&\cdots  & \textcolor{white}{00}1 \\
 \textcolor{white}{00} 0&\cdots  & \textcolor{white}{00}0 & \textcolor{white}{00}0 &\cdots &\textcolor{white}{00} 0& \textcolor{white}{00}0&\cdots  &\textcolor{white}{00} 5 \\
  a_{1,1}&\cdots  &   a_{1,6} & a_{1,7} &\cdots & a_{1,12}& a_{1,13} &\cdots  & a_{1,18} \\
    a_{2,1}&\cdots  &   a_{2,6} & a_{2,7} &\cdots & a_{2,12}& a_{2,13} &\cdots  & a_{2,18} \\
      a_{3,1}&\cdots  &   a_{3,6} & a_{3,7} &\cdots & a_{3,12}& a_{3,13} &\cdots  & a_{3,18} \\
        a_{4,1}&\cdots  &   a_{4,6} \tikzmark{right1} & a_{4,7}   &\cdots & a_{4,12} \tikzmark{right2} & a_{4,13} &\cdots  & a_{4,18}\tikzmark{right3} \ \ 
\end{bmatrix}. \label{eq:example F prime}
\end{equation}
\DrawBox[thick, black,   dashed]{left1}{right1}{\textcolor{black}{\footnotesize${\bf F^{\prime}}_1$}}
\DrawBox[thick, red, dashed]{left2}{right2}{\textcolor{red}{\footnotesize${\bf F^{\prime}}_2$}}
\DrawBox[thick, blue, dashed]{left3}{right3}{\textcolor{blue}{\footnotesize${\bf F^{\prime}}_3$}}

The transmissions of the $6$ workers can be expressed as 
\begin{align}
& {\bf S} \ {\bf F^{\prime}} \  [W_{1,1}; \ldots ; W_{6,1} ;W_{1,2}; \ldots; W_{6,3}] = 
[ \sv^{1,1}; \sv^{1,2};\sv^{2,1};\ldots; \sv^{6,2}] \ {\bf F^{\prime}} \ [W_{1,1}; \ldots ; W_{6,1} ;W_{1,2}; \ldots; W_{6,3}],
\end{align}
where the row vector $\sv^{n,j}$ represents the $j^{\text{th}}$ transmission vector  of worker $n$; in other words,
    $\sv^{n,j} {\bf F^{\prime}} [W_{1,1}; \ldots ; W_{6,1} ;W_{1,2}; \ldots; W_{6,3}] $     represents the $j^{\text{th}}$ transmitted linear combination by worker $n$. We can further expand ${\bf S}$ as follows, 
\begin{equation}\setstretch{1.25}
 {\bf S}= \left[ \begin{array}{c}
\sv^{1,1}\\
\sv^{1,2}\\
\sv^{2,1}\\
\sv^{2,2}\\
\vdots\\
\sv^{6,2}
\end{array} \right]= \begin{bmatrix}\ 
\tikzmark{left4} s^{1,1}_1 & s^{1,1}_2 & \tikzmark{left5} s^{1,1}_3 & s^{1,1}_4 &  \tikzmark{left6} s^{1,1}_5  & s^{1,1}_6  &  \tikzmark{left7} b^{1,1}_1  &  b^{1,1}_2 & b^{1,1}_3 & b^{1,1}_{4}   \\
s^{1,2}_1 & s^{1,2}_2 & s^{1,2}_3 & s^{1,2}_4 & s^{1,2}_5  & s^{1,2}_6  & b^{1,2}_1  &  b^{1,2}_2 & b^{1,2}_3 & b^{1,2}_{4}   \\
s^{2,1}_1 & s^{2,1}_2 & s^{2,1}_3 & s^{2,1}_4 & s^{2,1}_5  & s^{2,1}_6  & b^{2,1}_1  &  b^{2,1}_2 & b^{2,1}_3 & b^{2,1}_{4}   \\
\vdots &\vdots &\vdots &\vdots &\vdots &\vdots &\vdots &\vdots &\vdots &\vdots  \\ 
s^{6,2}_1 & s^{6,2}_2 \tikzmark{right4} & s^{6,2}_3 & s^{6,2}_4 \tikzmark{right5}  & s^{6,2}_5  & s^{6,2}_6 \tikzmark{right6} & b^{6,2}_1  &  b^{6,2}_2  & b^{6,2}_3 & b^{6,2}_{4} \tikzmark{right7} \ \
\end{bmatrix}. \label{eq:example division of S}
\end{equation}
 \DrawBox[thick, black,  dashed ]{left4}{right4}{\textcolor{black}{\footnotesize${\bf S}_1$}}
\DrawBox[thick, red, dashed]{left5}{right5}{\textcolor{red}{\footnotesize${\bf S}_2$}}
\DrawBox[thick, blue, dashed]{left6}{right6}{\textcolor{blue}{\footnotesize${\bf S}_3$}}
\DrawBox[thick, magenta, dashed]{left7}{right7}{\textcolor{magenta}{\footnotesize${\bf S}_4$}}

Now the   $j^{\text{th}}$ transmitted linear combination by worker $n$ can be expressed as   
 \begin{align}
 \sv^{n,j} \dv_{1} W_{1,1}+    \sv^{n,j} \dv_{2} W_{2,1} + \cdots+ \sv^{n,j} \dv_{6} W_{6,1}+ \sv^{n,j} \dv_{7} W_{1,2} +\cdots + \sv^{n,j} \dv_{18} W_{6,3}, \label{eq:example single sent}
 \end{align}
where $\dv_i$ represents the $i^{\text{th}}$ column of ${\bf F^{\prime}}$. Recall that $\overline{  \Zc_{n}} \subseteq [\Ksf]$ represents the set of messages which are not assigned to worker $n$. 
Hence, to guarantee that  the linear combination in~\eqref{eq:example single sent}  can be transmitted by worker $n$, we should have 
\begin{align}
\sv^{n,j} \dv_{k+(t-1)\Ksf} =0, \  \forall n\in [6], j\in [2], t\in [3], k\in \overline{  \Zc_{n}}.  \label{eq:example transmit constraint}
\end{align} 
In addition, for each set $\Ac \subseteq [6]$ where $|\Ac|=5$, by receiving the linear combinations transmitted by the workers in $\Ac$, the master should recover the desired linear combinations. Hence, we should have (recalling that $\Ac(i)$ represents the $i^{\text{th}}$ smallest element of $\Ac$)
\begin{align}
  [\sv^{\Ac(1),1};\sv^{\Ac(1),2};\sv^{\Ac(2),1};\ldots;\sv^{\Ac(5),2}]  
   \ \text{is full rank}, \  \forall \Ac\subseteq [6]: |\Ac|=5. \label{eq:example decodability constraint}
\end{align}
Our objective is to determine the variables in ${\bf S}$ and in ${\bf F^{\prime}}$ such that the constraints in~\eqref{eq:example transmit constraint} and~\eqref{eq:example decodability constraint} are satisfied. 

We divide matrix ${\bf F^{\prime}}$ into $3$ sub-matrices,  ${\bf F^{\prime}}_1, {\bf F^{\prime}}_2, {\bf F^{\prime}}_3$  each of which has the dimension $10 \times 6$, as illustrated in~\eqref{eq:example F prime}. We also    divide matrix ${\bf S}$ into $4$ sub-matrices, ${\bf S}_1, {\bf S}_2, {\bf S}_3$  each of which has the dimension $12 \times 2$ and ${\bf S}_4$ with dimension $12 \times 4$, as illustrated in~\eqref{eq:example division of S}.

The proposed computing scheme in the computing phase contains three main steps:\footnote{\label{foot:cannot use}  Notice that the computing schemes in~\cite{efficientgradientcoding,adaptiveGC2020} for the case  $\Ksf_{\rm c}=1$ and in~\cite{linearcomput2020wan} for the case where $\msf=1$ cannot be used in this example to achieved the converse bound. The idea of the computing schemes in~\cite{efficientgradientcoding,adaptiveGC2020} is   first to randomly determine the variables in ${\bf S}$, and then to determine the coefficients of the virtually demanded linear combinations in ${\bf F^{\prime}}$ in order to satisfy the constraints in~\eqref{eq:example transmit constraint}. One can check that if we randomly choose all the variables in ${\bf S}$, there does not exist any solution on ${\bf F^{\prime}}$ which satisfies the constraints in~\eqref{eq:example transmit constraint}, because there will be more linearly independent constraints than the variables. The idea of the computing scheme in~\cite{linearcomput2020wan} is  first to randomly determine the coefficients of the virtually demanded linear combinations in  ${\bf F^{\prime}}$, and then to determine the  variables in ${\bf S}$ in order to satisfy the constraints in~\eqref{eq:example transmit constraint}. However, one can check that if we randomly determine the coefficients of the virtually demanded linear combinations in  ${\bf F^{\prime}}$, we cannot find any solution of  ${\bf S}$    satisfying the constraints in~\eqref{eq:example transmit constraint}, where the two transmission vectors of each worker in ${\bf S}$  are linearly independent.
}
\begin{enumerate}
\item we first   choose the values for the variables in ${\bf S}_4$;
\item after determining   ${\bf S}_4$, the constraints in~\eqref{eq:example single sent} become  linear in terms of the remaining variables (i.e., the variables in ${\bf F^{\prime}}_1, {\bf F^{\prime}}_2, {\bf F^{\prime}}_3, {\bf S}_1, {\bf S}_2, {\bf S}_3$). Hence, we can obtain the values for these remaining variables by solving linear equations;
\item after determining all the variables, we check that the constraints in~\eqref{eq:example decodability constraint} such that the proposed scheme is decodable. 
\end{enumerate}

\paragraph*{Step 1}
We choose the values for ${\bf S}_4$ with the following form,
\begin{align}
{\bf S}_4= \begin{bmatrix}\ 
  b^{1,1}_1  &  b^{1,1}_2 & b^{1,1}_3 & b^{1,1}_{4}   \\
  b^{1,2}_1  &  b^{1,2}_2 & b^{1,2}_3 & b^{1,2}_{4}   \\
  b^{2,1}_1  &  b^{2,1}_2 & b^{2,1}_3 & b^{2,1}_{4}   \\
  b^{2,2}_1  &  b^{2,2}_2 & b^{2,2}_3 & b^{2,2}_{4}   \\
    b^{3,1}_1  &  b^{3,1}_2 & b^{3,1}_3 & b^{3,1}_{4}   \\
  b^{3,2}_1  &  b^{3,2}_2 & b^{3,2}_3 & b^{3,2}_{4}   \\
  b^{4,1}_1  &  b^{4,1}_2 & b^{4,1}_3 & b^{4,1}_{4}   \\
  b^{4,2}_1  &  b^{4,2}_2 & b^{4,2}_3 & b^{4,2}_{4}   \\
    b^{5,1}_1  &  b^{5,1}_2 & b^{5,1}_3 & b^{5,1}_{4}   \\
  b^{5,2}_1  &  b^{5,2}_2 & b^{5,2}_3 & b^{5,2}_{4}   \\
  b^{6,1}_1  &  b^{6,1}_2 & b^{6,1}_3 & b^{6,1}_{4}   \\
  b^{6,2}_1  &  b^{6,2}_2 & b^{6,2}_3 & b^{6,2}_{4} 
\end{bmatrix} 
=
 \begin{bmatrix}  
  * &  * & 0 & 0   \\
  0  &  0 & * & *   \\
  * &  * & 0 & 0   \\
  0  &  0 & * & *   \\
    * &  * & 0 & 0   \\
  0  &  0 & * & *   \\
    * &  * & 0 & 0   \\
  0  &  0 & * & *   \\
    * &  * & 0 & 0   \\
  0  &  0 & * & *   \\
    * &  * & 0 & 0   \\
  0  &  0 & * & *   
\end{bmatrix}
=
 \begin{bmatrix}  
  0 &  2 & 0 & 0   \\
  0  &  0 & 2 & 0    \\
  2 & 2 & 0 & 0   \\
  0  &  0 & 0 & 2   \\
   1 & 2 & 0 & 0   \\
  0  &  0 & 2 & 1 \\
   0 & 1 & 0 & 0   \\
  0  &  0 & 1 & 0   \\
    1&0 & 0 & 0   \\
  0  &  0 & 2&1   \\
    2& 2 & 0 & 0   \\
  0  &  0 & 1&1   
\end{bmatrix},\label{eq:example S4 assignment}
\end{align}
where each `$*$' represents an  uniform i.i.d.    symbol   on $\mathbb{F}_{\qsf}$. 
More precisely,  for the first linear combination transmitted by each worker $n\in [6]$, we choose $b^{n,1}_{1}$ and  $b^{n,1}_{2}$ uniformly i.i.d. over   $\mathbb{F}_{\qsf}$, while letting  $b^{n,1}_{3}$ and  $b^{n,1}_{4}$ be zero. For the second linear combination transmitted by each worker $n$, we choose $b^{n,2}_{3}$ and  $b^{n,2}_{4}$ uniformly i.i.d. over $\mathbb{F}_{\qsf}$, while letting  $b^{n,2}_{1}$ and  $b^{n,2}_{2}$ be zero.
The above choice on ${\bf S}_4$ will guarantee that 
the  constraints in~\eqref{eq:example single sent} become  linearly independent  in terms of the remaining variables to be decided in the next step.

\paragraph*{Step 2}
Let us  focus on the constraints in~\eqref{eq:example transmit constraint} for $t=1$, which corresponds to the variables in ${\bf S}_1$ and ${\bf F^{\prime}}_1$.

When $(t,j)=(1,1)$,  the constraints in~\eqref{eq:example transmit constraint}  become
\begin{align}
s^{n,1}_1 f_{1,k} + s^{n,1}_2 f_{2,k} + b^{n,1}_1 a_{1,k} + b^{n,1}_2 a_{2,k} +b^{n,1}_3 a_{3,k} +b^{n,1}_{4} a_{4,k}  =0, \  \forall n\in [6],   k\in \overline{  \Zc_{n}},
\end{align}
where $f_{1,k}$ represents the $k^{\text{th}}$ element in the first   demand vector, 
 $f_{2,k}$ represents the $k^{\text{th}}$ element in the second   demand vector,
and the values of $b^{n,1}_{i}$ where $i\in [4]$ have been chosen in~\eqref{eq:example S4 assignment}. 
For example, if $n=1$, we have the set of datasets which are not assigned to worker $1$ is $ \overline{  \Zc_{1}}=\{ 4,5,6\}$. Hence, we have the following three constraints
   \begin{subequations}
\begin{align}
& s^{1,1}_1 f_{1,4} + s^{1,1}_2 f_{2,4} + b^{1,1}_1 a_{1,4} + b^{1,1}_2 a_{2,4} +b^{1,1}_3 a_{3,4} +b^{1,1}_{4} a_{4,4}  = 1 s^{1,1}_1 + 3 s^{1,1}_2 + 0  a_{1,4}+ 2 a_{2,4} =0, \nonumber\\ 
& s^{1,1}_1 f_{1,5} + s^{1,1}_2 f_{2,5} + b^{1,1}_1 a_{1,5} + b^{1,1}_2 a_{2,5} +b^{1,1}_3 a_{3,5} +b^{1,1}_{4} a_{4,5}  = 1 s^{1,1}_1 + 4 s^{1,1}_2 + 0  a_{1,5}+ 2 a_{2,5} =0, \nonumber\\ 
& s^{1,1}_1 f_{1,6} + s^{1,1}_2 f_{2,6} + b^{1,1}_1 a_{1,6} + b^{1,1}_2 a_{2,6} +b^{1,1}_3 a_{3,6} +b^{1,1}_{4} a_{4,6}  = 1 s^{1,1}_1 + 5 s^{1,1}_2 + 0  a_{1,6}+ 2 a_{2,6} =0. \nonumber  
\end{align} 
   \end{subequations}
Similarly, if $n=2$, with $ \overline{  \Zc_{2}}=\{ 1,5,6\}$  we have the following three constraints
   \begin{subequations}
\begin{align}
& s^{2,1}_1 f_{1,1} + s^{2,1}_2 f_{2,1} + b^{2,1}_1 a_{1,1} + b^{2,1}_2 a_{2,1} +b^{2,1}_3 a_{3,1} +b^{2,1}_{4} a_{4,1}  =  1 s^{2,1}_1 + 0 s^{2,1}_2 + 2  a_{1,1}+ 2 a_{2,1} =0,\nonumber\\ 
& s^{2,1}_1 f_{1,5} + s^{2,1}_2 f_{2,5} + b^{2,1}_1 a_{1,5} + b^{2,1}_2 a_{2,5} +b^{2,1}_3 a_{3,5} +b^{2,1}_{4} a_{4,5}  = 1 s^{2,1}_1 + 4 s^{2,1}_2 + 2  a_{1,5}+ 2 a_{2,5} =0, \nonumber\\ 
& s^{2,1}_1 f_{1,6} + s^{2,1}_2 f_{2,6} + b^{2,1}_1 a_{1,6} + b^{2,1}_2 a_{2,6} +b^{2,1}_3 a_{3,6} +b^{2,1}_{4} a_{4,6}  = 1 s^{2,1}_1 + 5 s^{2,1}_2 + 2  a_{1,6}+ 2 a_{2,6} =0. \nonumber  
\end{align}
   \end{subequations}
If $n=3$, with $ \overline{  \Zc_{3}}=\{ 1,2,6\}$  we have the following three constraints
   \begin{subequations}
\begin{align}
& s^{3,1}_1 f_{1,1} + s^{3,1}_2 f_{2,1} + b^{3,1}_1 a_{1,1} + b^{3,1}_2 a_{2,1} +b^{3,1}_3 a_{3,1} +b^{3,1}_{4} a_{4,1}  = 1 s^{3,1}_1 + 0 s^{3,1}_2 + 1  a_{1,1}+ 2 a_{2,1} =0, \nonumber\\ 
& s^{3,1}_1 f_{1,2} + s^{3,1}_2 f_{2,2} + b^{3,1}_1 a_{1,2} + b^{3,1}_2 a_{2,2} +b^{3,1}_3 a_{3,2} +b^{3,1}_{4} a_{4,2}  =  1 s^{3,1}_1 + 1 s^{3,1}_2 + 1  a_{1,2}+ 2 a_{2,2} =0, \nonumber\\ 
& s^{3,1}_1 f_{1,6} + s^{3,1}_2 f_{2,6} + b^{3,1}_1 a_{1,6} + b^{3,1}_2 a_{2,6} +b^{3,1}_3 a_{3,6} +b^{3,1}_{4} a_{4,6}  = 1 s^{3,1}_1 + 5 s^{3,1}_2 + 1  a_{1,6}+ 2 a_{2,6} =0.  \nonumber  
\end{align}
   \end{subequations}
If $n=4$, with $ \overline{  \Zc_{4}}=\{ 1,2,3\}$  we have the following three constraints
   \begin{subequations}
\begin{align}
& s^{4,1}_1 f_{1,1} + s^{4,1}_2 f_{2,1} + b^{4,1}_1 a_{1,1} + b^{4,1}_2 a_{2,1} +b^{4,1}_3 a_{3,1} +b^{4,1}_{4} a_{4,1}  = 1 s^{4,1}_1 + 0 s^{4,1}_2 + 0  a_{1,1}+ 1 a_{2,1} =0, \nonumber\\ 
& s^{4,1}_1 f_{1,2} + s^{4,1}_2 f_{2,2} + b^{4,1}_1 a_{1,2} + b^{4,1}_2 a_{2,2} +b^{4,1}_3 a_{3,2} +b^{4,1}_{4} a_{4,2}  =  1 s^{4,1}_1 + 1 s^{4,1}_2 + 0  a_{1,2}+ 1 a_{2,2} =0, \nonumber\\ 
& s^{4,1}_1 f_{1,3} + s^{4,1}_2 f_{2,3} + b^{4,1}_1 a_{1,3} + b^{4,1}_2 a_{2,3} +b^{4,1}_3 a_{3,3} +b^{4,1}_{4} a_{4,3}  = 1 s^{4,1}_1 + 2 s^{4,1}_2 + 0  a_{1,3}+ 1 a_{2,3} =0. \nonumber 
\end{align}
   \end{subequations}
If $n=5$, with $ \overline{  \Zc_{5}}=\{2,3,4\}$  we have the following three constraints
   \begin{subequations}
\begin{align}
& s^{5,1}_1 f_{1,2} + s^{5,1}_2 f_{2,2} + b^{5,1}_1 a_{1,2} + b^{5,1}_2 a_{2,2} +b^{5,1}_3 a_{3,2} +b^{5,1}_{4} a_{4,2}  = 1 s^{5,1}_1 + 1 s^{5,1}_2 + 1  a_{1,2}+ 0 a_{2,2} =0,  \nonumber\\ 
& s^{5,1}_1 f_{1,3} + s^{5,1}_2 f_{2,3} + b^{5,1}_1 a_{1,3} + b^{5,1}_2 a_{2,3} +b^{5,1}_3 a_{3,3} +b^{5,1}_{4} a_{4,3}  =  1 s^{5,1}_1 + 2 s^{5,1}_2 + 1  a_{1,3}+ 0 a_{2,3} =0,\nonumber\\ 
& s^{5,1}_1 f_{1,4} + s^{5,1}_2 f_{2,4} + b^{5,1}_1 a_{1,4} + b^{5,1}_2 a_{2,4} +b^{5,1}_3 a_{3,4} +b^{5,1}_{4} a_{4,4}  = 1 s^{5,1}_1 + 3 s^{5,1}_2 + 1  a_{1,4}+ 0 a_{2,4} =0. \nonumber  
\end{align}
   \end{subequations}
If $n=6$, with $ \overline{  \Zc_{6}}=\{3,4,5\}$  we have the following three constraints
   \begin{subequations}
\begin{align}
& s^{6,1}_1 f_{1,3} + s^{6,1}_2 f_{2,3} + b^{6,1}_1 a_{1,3} + b^{6,1}_2 a_{2,3} +b^{6,1}_3 a_{3,3} +b^{6,1}_{4} a_{4,3}  = 1 s^{6,1}_1 + 2 s^{6,1}_2 + 2  a_{1,3}+ 2 a_{2,3} =0,\nonumber\\ 
& s^{6,1}_1 f_{1,4} + s^{6,1}_2 f_{2,4} + b^{6,1}_1 a_{1,4} + b^{6,1}_2 a_{2,4} +b^{6,1}_3 a_{3,4} +b^{6,1}_{4} a_{4,4}  =  1 s^{6,1}_1 + 3 s^{6,1}_2 + 2  a_{1,4}+ 2 a_{2,4} =0,\nonumber\\ 
& s^{6,1}_1 f_{1,5} + s^{6,1}_2 f_{2,5} + b^{6,1}_1 a_{1,5} + b^{6,1}_2 a_{2,5} +b^{6,1}_3 a_{3,5} +b^{6,1}_{4} a_{4,5}  = 1 s^{6,1}_1 + 4 s^{6,1}_2 + 2  a_{1,5}+ 2 a_{2,5} =0. \nonumber  
\end{align}
   \end{subequations}
Hence,  there are totally $6\times 3=18$ constraints on $24$ variables, which are 
\begin{align}
a_{1,1},\ldots,a_{1,6}, a_{2,1},\ldots,a_{2,6},s^{1,1}_1,s^{1,1}_2,s^{2,1}_1,s^{2,1}_2,\ldots,s^{6,1}_2. \label{eq:example tj1 all variables}
\end{align}
We then give a random value to each of $s^{1,1}_1, s^{2,1}_2,s^{3,1}_1,s^{4,1}_2,s^{5,1}_1,s^{6,1}_1$, totally $6$ variables among the $24$ variables in~\eqref{eq:example tj1 all variables}, as follows, 
\begin{align}
s^{1,1}_1=0, \  s^{2,1}_2=1 , \ s^{3,1}_1=1, \ s^{4,1}_2=1, \ s^{5,1}_1=0, \ s^{6,1}_2 =1. \label{eq:example tj1 first 6 variables}         
\end{align}  
After determining the $6$ variables in~\eqref{eq:example tj1 first 6 variables}, it can be checked that the above $18$ constraints 
are linearly independent on the remaining $18$ variables, such that we can solve 
   \begin{subequations}
\begin{align}
&a_{1,1}= 1/4, \ a_{1,2}=5/8, \ a_{1,3}=5/4, \ a_{1,4}=15/8, \ a_{1,5}=21/8, \ a_{1,6}=27/8, \\
&a_{2,1}=-5/8, \ a_{2,2}=-13/8, \ a_{2,3}= -21/8, \ a_{2,4}=-15/4, \ a_{2,5}=-5, \ a_{2,6}=-25/4,\\
&s^{1,1}_2= 5/2, \  s^{2,1}_1=3/4  , \ s^{3,1}_2= 13/8, \ s^{4,1}_1= 5/8, \ s^{5,1}_2= - 5/8 , \ s^{6,1}_1 = 3/4.
\end{align}
\label{eq:example tj1 other 18 variables}  
   \end{subequations}

Similarly, by  considering all pairs $(t,j)$ where $t\in [3]$ and $j\in [2]$,
we can determine~\eqref{eq:example solution}. 
    \begin{figure*}
   \begin{subequations}
\begin{align}
& {\bf S}= \begin{bmatrix}  
  0& 5/2 & 0 & 0 & 0 & - 11/4 & 0 & 2 & 0& 0   \\
1& -14& 1& 27& 0& 0& 0& 0& 2& 0  \\
3/4&  1 & 0& 0& 41/8& 1& 2& 2& 0& 0  \\
40& 0& -82& 1& 0& 0& 0& 0& 0& 2  \\ 
1& 13/8& 0& 0& 1& - 9/16 & 1&  2& 0& 0 \\
1& -10& 0& 39/2& 0& 0& 0& 0& 2& 1 \\
 5/8& 1& 0& 0& - 25/16&  0& 0& 1& 0& 0 \\
 - 19/2 & 0& 41/2& 1& 0& 0& 0& 0& 1& 0 \\
 0& - 5/8&  0& 0& 1& 41/16& 1& 0& 0& 0 \\
 1& -10& 1& 37/2& 0& 0& 0& 0& 2& 1 \\
 3/4& 1& 0& 0& 73/8& 0& 2& 2& 0& 0 \\
 - 23/2& 1& 31/2& 0& 0& 0& 0& 0& 1& 1  
\end{bmatrix} ;   \\
&
[a_{1,1}, \ldots, a_{1,18}]=  \left[\frac{1}{4},  \frac{5}{8}  , \frac{5}{4} , \frac{15}{8}, \frac{21}{8} , \frac{27}{8}, 0,  0,  0,  0,  0,  0, \frac{-33}{8}, \frac{-57}{16} , \frac{-49}{8}, \frac{139}{16}, \frac{161}{16}, \frac{-191}{16} \right]; \\
& [a_{2,1}, \ldots, a_{2,18}]= \left[\frac{-5}{8}, \frac{-13}{8}, \frac{-21}{8}, \frac{-15}{4},-5,\frac{-25}{4},  0,  0,  0,  0,  0,  0, \frac{25}{16}, \frac{25}{16}, \frac{25}{16}, \frac{33}{8}, \frac{11}{2}, \frac{55}{8}  \right]; \\
& [a_{3,1}, \ldots, a_{3,18}]= \left[\frac{19}{2}   ,  \frac{19}{2} , \frac{19}{2} , \frac{41}{2} , \frac{55}{2}   , \frac{69}{2}    , \frac{-41}{2}      , \frac{-43}{2}    , \frac{-45}{2}      ,  -41  ,  \frac{-109}{2}    , -68 , 0,  0,  0,  0,  0,  0 \right]; \\
& [a_{4,1}, \ldots, a_{4,18}]= \left[ -20 , -10, 0, -12 , -20 , -20 , 41  , \frac{47}{2}  ,  7 ,  \frac{51}{2}  ,  39  , \frac{77}{2}  ,0,  0,  0,  0,  0,  0 \right].
\end{align}
\label{eq:example solution}
   \end{subequations}
     \end{figure*}

\paragraph*{Step 3}
For each subset of workers $\Ac\subseteq [6]$ where $|\Ac|=5$, it can be seen that the constraints in~\eqref{eq:example decodability constraint} holds. For example, if $\Ac=[5]$, the sub-matrix ${\bf S}^{([10])_{\rm r}}$ including the first $10$ rows of ${\bf S}$ is full-rank. Hence, we let each worker $n$ compute and send  two linear combinations of sub-messages, $\sv^{n,1} {\bf F^{\prime}} [W_{1,1};\ldots; W_{6,3}]$ and $\sv^{n,2} {\bf F^{\prime}} [W_{1,1};\ldots; W_{6,3}]$.

\paragraph*{Decoding phase}
Assume that the set of responding workers is $\Ac$ where $\Ac\subseteq [6]$ and $|\Ac|=5$. The master receives 
\begin{align}
{\bf X}_{\Ac}=  \left[ \sv^{\Ac(1),1}; \sv^{\Ac(1),2}; \sv^{\Ac(2),1}; \ldots ; \sv^{\Ac(5),2}\right]  {\bf F^{\prime}} \  [W_{1,1}; \ldots ; W_{6,1} ;W_{1,2}; \ldots; W_{6,3}].
\end{align}
Since $ \left[ \sv^{\Ac(1),1}; \sv^{\Ac(1),2}; \sv^{\Ac(2),1}; \ldots ; \sv^{\Ac(5),2}\right] $ is full-rank, the master then computes
$$
 \left[ \sv^{\Ac(1),1}; \sv^{\Ac(1),2}; \sv^{\Ac(2),1}; \ldots ; \sv^{\Ac(5),2}\right] ^{-1} {\bf X}_{\Ac}
$$
to obtain $ {\bf F^{\prime}} \  [W_{1,1}; \ldots ; W_{6,1} ;W_{1,2}; \ldots; W_{6,3}]$, which  contains its demanded linear combinations.

 \paragraph*{Performance}
Since each worker sends $\frac{2 \Lsf}{3}$ symbols, the   communication cost is $\frac{10 \Lsf}{3 \Lsf}= \frac{10}{3}$, coinciding with the converse bound in~\eqref{eq:case 2 converse}.
 \hfill $\square$ 
\end{example}

We are ready to generalize the proposed distributed computing scheme in Example~\ref{ex:achie example n=k first reg}. First we focus on  $\Ksf_{\rm c}= \frac{\Ksf}{\Nsf} \usf$, where $\usf \in [\Nsf_{\rm r}-\msf+1]$ and $\Nsf \geq \frac{\msf+\usf-1}{\usf}+ \usf(\Nsf_{\rm r}-\msf-\usf+1)$. 
During the data assignment phase, we use the cyclic assignment.

\paragraph*{Computing phase}
Since the     communication cost is no less than  
$
\Nsf_{\rm r} \frac{\Ksf_{\rm c}}{\msf+\usf-1},
$ 
from the converse bound~\eqref{eq:case 2 converse}, 
we divide each message $W_k$ where $k\in [\Ksf]$ into $\msf+\usf-1$ non-overlapping and equal-length sub-messages,
$W_k=\{W_{k,j}:j\in [\msf+\usf-1]\}.$ 
 Each worker should send $\Ksf_{\rm c}$  linear combinations of sub-messages. 
From the answers of $\Nsf_{\rm r}$  workers, the master totally receives $ \Nsf_{\rm r} \Ksf_{\rm c}$ linear combinations of sub-messages. Hence, 
we generate 
$$
\vsf = \Nsf_{\rm r} \Ksf_{\rm c} - (\msf+\usf-1)\Ksf_{\rm c} =\Ksf_{\rm c} (\Nsf_{\rm r}-\msf-\usf+1)
$$
 virtually requested linear combinations; thus the effective demand matrix  ${\bf F^{\prime}}  $ has dimension $ \Nsf_{\rm r}\Ksf_{\rm c}  \times \Ksf(\msf+\usf-1) $,  
 with the form   in~\eqref{eq:general F prime}.
 
    \begin{figure*}
\begin{equation}\setstretch{1.25}
{\bf F^{\prime}} =\begin{bmatrix}  \     
\tikzmark{left1} \textcolor{white}{0} f_{1,1} &\cdots &  \textcolor{white}{0} f_{1,\Ksf}& \tikzmark{left2} \textcolor{white}{000} 0&\cdots  & \textcolor{white}{000} 0& \cdots & \tikzmark{left3}  \textcolor{white}{00000} 0 \textcolor{white}{00000} & \cdots &  \textcolor{white}{00000} 0 \textcolor{white}{00000} \\
\vdots &\ddots & \vdots &  \vdots & \ddots  & \vdots & \ddots & \vdots  & \ddots & \vdots\\
 f_{\Ksf_{\rm c},1} &\cdots & f_{\Ksf_{\rm c},\Ksf}&  \textcolor{white}{00} 0&\cdots  & \textcolor{white}{00} 0& \cdots &   \textcolor{white}{00000} 0 \textcolor{white}{00000} & \cdots & \textcolor{white}{00000} 0 \textcolor{white}{00000}\\
\textcolor{white}{00}0  &\cdots &\textcolor{white}{00}0  &  f_{1,1} &\cdots  & f_{1,\Ksf} & \cdots &   \textcolor{white}{00000} 0 \textcolor{white}{00000}& \cdots & \textcolor{white}{00000} 0 \textcolor{white}{00000} \\
\vdots &\ddots & \vdots &  \vdots & \ddots  & \vdots & \ddots & \vdots  & \ddots & \vdots\\
\textcolor{white}{00}0  &\cdots &\textcolor{white}{00}0  &  f_{\Ksf_{\rm c},1} &\cdots  & f_{\Ksf_{\rm c},\Ksf} & \cdots &   \textcolor{white}{00000} 0 \textcolor{white}{00000} & \cdots &\textcolor{white}{00000} 0 \textcolor{white}{00000}\\
\vdots &\ddots & \vdots &  \vdots & \ddots  & \vdots & \ddots & \vdots  & \ddots & \vdots\\
 \textcolor{white}{00}0  &\cdots &\textcolor{white}{00}0  &  \textcolor{white}{00}0 &\cdots  & \textcolor{white}{00}0 & \cdots &  f_{1,1}  & \cdots &  f_{1,\Ksf} \\
\vdots &\ddots & \vdots &  \vdots & \ddots  & \vdots & \ddots & \vdots  & \ddots & \vdots\\
 \textcolor{white}{00}0  &\cdots &\textcolor{white}{00}0  &  \textcolor{white}{00}0 &\cdots  & \textcolor{white}{00}0 & \cdots &  f_{\Ksf_{\rm c},1}  & \cdots &  f_{\Ksf_{\rm c},\Ksf} \\
  a_{1,1}&\cdots  &   a_{1,\Ksf} & a_{1,\Ksf+1} &\cdots & a_{1,2\Ksf}& \cdots &  a_{1,(\msf+\usf-2)\Ksf+1 } &\cdots  & a_{1, (\msf+\usf-1)\Ksf} \\
\vdots &\ddots & \vdots &  \vdots & \ddots  & \vdots & \ddots & \vdots  & \ddots & \vdots\\
    a_{\vsf,1}&\cdots  &   a_{\vsf,\Ksf} \tikzmark{right1}  & a_{\vsf ,\Ksf+1} &\cdots & a_{\vsf,2\Ksf} \tikzmark{right2}  & \cdots &  a_{\vsf,(\msf+\usf-2)\Ksf+1 } &\cdots  & a_{\vsf, (\msf+\usf-1)\Ksf} \tikzmark{right3}  \ \ 
\end{bmatrix}. \label{eq:general F prime}
\end{equation}
\DrawBox[thick, black,   dashed]{left1}{right1}{\textcolor{black}{\footnotesize${\bf F^{\prime}}_1$}}
\DrawBox[thick, red, dashed]{left2}{right2}{\textcolor{red}{\footnotesize${\bf F^{\prime}}_2$}}
\DrawBox[thick, blue, dashed]{left3}{right3}{\textcolor{blue}{\footnotesize${\bf F^{\prime}}_{\msf+\usf-1}$}}
   \end{figure*}

The transmissions of the $\Ksf$ workers can be expressed as 
\begin{align}
& {\bf S} \ {\bf F^{\prime}}  \   [W_{1,1}; \ldots ; W_{\Ksf,1} ;W_{1,2}; \ldots; W_{\Ksf,\msf+\usf-1}] \nonumber\\
& = 
[\sv^{1,1};\ldots;\sv^{1,\Ksf_{\rm c}};\sv^{2,1};\ldots;\sv^{\Nsf,\Ksf_{\rm c}}] \ {\bf F^{\prime}} \   [W_{1,1}; \ldots ; W_{\Ksf,1} ;W_{1,2}; \ldots; W_{\Ksf,\msf+\usf-1}],
\end{align}
where     $\sv^{n,j} {\bf F^{\prime}}[W_{1,1}; \ldots ; W_{\Ksf,1} ;W_{1,2}; \ldots; W_{\Ksf,\msf+\usf-1}] $     represents the $j^{\text{th}}$ transmitted linear combination by worker $n$. We can further expand ${\bf S}$ as follows, 
\begin{equation}\setstretch{1.25}
 {\bf S}=  \left[ \begin{array}{c}
\sv^{1,1}\\
\vdots\\
\sv^{1,\Ksf_{\rm c}}\\
\sv^{2,1}\\
\vdots\\
\sv^{\Nsf,\Ksf_{\rm c}}
\end{array} \right] =  
\begin{bmatrix} \ 
\tikzmark{left4} \textcolor{white}{0} s^{1,1}_1 & \cdots& s^{1,1}_{\Ksf_{\rm c}} & \cdots & 
\tikzmark{left6} s^{1,1}_{(\msf+\usf-2)\Ksf_{\rm c}+1} &\cdots & s^{1,1}_{(\msf+\usf-1)\Ksf_{\rm c}}  &   
  \tikzmark{left7} \textcolor{white}{0} b^{1,1}_{1}  & \cdots&  b^{1,1}_{\vsf}   \\
\vdots & \ddots &  \vdots  & \ddots & \vdots & \ddots & \vdots & \vdots & \ddots &\vdots \\
  s^{1,\Ksf_{\rm c}}_1 & \cdots& s^{1,\Ksf_{\rm c}}_{\Ksf_{\rm c}} &   \cdots & 
 s^{1,\Ksf_{\rm c}}_{(\msf+\usf-2)\Ksf_{\rm c}+1} &\cdots & s^{1,\Ksf_{\rm c}}_{(\msf+\usf-1)\Ksf_{\rm c}}  &   
 b^{1,\Ksf_{\rm c}}_{1}  & \cdots& b^{1,\Ksf_{\rm c}}_{\vsf}   \\
   s^{2,1}_1 & \cdots& s^{2,1}_{\Ksf_{\rm c}} &   \cdots & 
 s^{2,1}_{(\msf+\usf-2)\Ksf_{\rm c}+1} &\cdots & s^{2,1}_{(\msf+\usf-1)\Ksf_{\rm c}}  &   
b^{2,1}_{1}  & \cdots&  b^{2,1}_{\vsf}   \\
\vdots & \ddots &  \vdots  & \ddots & \vdots & \ddots & \vdots & \vdots & \ddots &\vdots \\
 s^{\Nsf,\Ksf_{\rm c}}_1 & \cdots& s^{\Nsf,\Ksf_{\rm c}}_{\Ksf_{\rm c}} \tikzmark{right4} &      \cdots & 
 s^{\Nsf,\Ksf_{\rm c}}_{(\msf+\usf-2)\Ksf_{\rm c}+1} &\cdots & s^{\Nsf,\Ksf_{\rm c}}_{(\msf+\usf-1)\Ksf_{\rm c}} \tikzmark{right6}  &   
 b^{\Nsf,\Ksf_{\rm c}}_{1}  & \cdots&  b^{\Nsf,\Ksf_{\rm c}}_{\vsf} \tikzmark{right7}  \ \
 \end{bmatrix}. \label{eq:division of S}
\end{equation}
 \DrawBox[thick, black,  dashed ]{left4}{right4}{\textcolor{black}{\footnotesize${\bf S}_1$}}
\DrawBox[thick, blue, dashed]{left6}{right6}{\textcolor{blue}{\footnotesize${\bf S}_{\msf+\usf-1}$}}
\DrawBox[thick, magenta, dashed]{left7}{right7}{\textcolor{magenta}{\footnotesize${\bf S}_{\msf+\usf}$}}

By defining $\dv_i$ as the $i^{\text{th}}$ column of ${\bf F^{\prime}}$, the   $j^{\text{th}}$ transmitted linear combination by worker $n$ can be expressed as   
 \begin{align}
 \sv^{n,j} \dv_{1} W_{1,1}+     \cdots+ \sv^{n,j} \dv_{\Ksf} W_{\Ksf,1}+ \sv^{n,j} \dv_{\Ksf+1} W_{1,2} +\cdots + \sv^{n,j} \dv_{(\msf+\usf-1)\Ksf} W_{\Ksf,\msf+\usf-1}. \label{eq:single sent}
 \end{align}
To guarantee that  the linear combination in~\eqref{eq:example single sent}  can be transmitted by worker $n$,   the coefficients of the sub-messages which worker $n$ cannot compute should be $0$; that is 
\begin{align}
\sv^{n,j} \dv_{k+(t-1)\Ksf} =0, \  \forall n\in [\Nsf], j\in [\Ksf_{\rm c}], t\in [\msf+\usf-1], k\in \overline{  \Zc_{n}}.  \label{eq:transmit constraint}
\end{align} 
In addition, for each set $\Ac \subseteq [\Nsf]$ where $|\Ac|=\Nsf_{\rm r}$, by receiving the linear combinations transmitted by the workers in $\Ac$, the master should recover the desired linear combinations. Hence, we should have 
\begin{align}
[\sv^{\Ac(1),1};\ldots;\sv^{\Ac(1),\Ksf_{\rm c}};\sv^{\Ac(2),1};\ldots;\sv^{\Ac(\Nsf_{\rm r}),\Ksf_{\rm c}}]   \ \text{is full rank}, \  \forall \Ac\subseteq [\Nsf]: |\Ac|=\Nsf_{\rm r}. \label{eq:decodability constraint}
\end{align}
Our objective is to determine the variables in ${\bf S}$ (i.e., $s^{n,j}_{i}$ where $n\in [\Nsf]$, $j\in [\Ksf_{\rm c}]$, $i\in [(\msf+\usf-1)\Ksf_{\rm c}]$; $b^{n,j}_i$ where $n\in [\Nsf]$, $j\in [\Ksf_{\rm c}]$, $i\in [\vsf]$) and in ${\bf F^{\prime}}$ (i.e., $a_{i,k}$ where $i\in [\vsf]$ and $k\in [(\msf+\usf-1)\Ksf]$) such that the constraints in~\eqref{eq:transmit constraint} and~\eqref{eq:decodability constraint} are satisfied.

We divide matrix ${\bf F^{\prime}}$ into $\msf+\usf-1$ sub-matrices,  ${\bf F^{\prime}}_1, \ldots, {\bf F^{\prime}}_{\msf+\usf-1}$  each of which has the dimension $\Nsf_{\rm r}\Ksf_{\rm c} \times \Ksf$, as illustrated in~\eqref{eq:general F prime}. We also    divide matrix ${\bf S}$ into $\msf+\usf$ sub-matrices, ${\bf S}_1, \ldots, {\bf S}_{\msf+\usf-1}$  each of which has the dimension $\Nsf \Ksf_{\rm c} \times \Ksf_{\rm c}$ and ${\bf S}_{\msf+\usf}$ with dimension $\Nsf \Ksf_{\rm c} \times \vsf$, as illustrated in~\eqref{eq:division of S}.
As in Example~\ref{ex:achie example n=k first reg}, the proposed computing scheme contains three  main steps:
\begin{enumerate}
\item we first   choose the values for the variables in ${\bf S}_{\msf+\usf}$;
\item after determining the variables in ${\bf S}_{\msf+\usf}$, the constraints in~\eqref{eq:transmit constraint} become   linear in terms of the remaining variables, which are then determined  by solving linear equations;
\item after determining all the variables, we check that the constraints in~\eqref{eq:decodability constraint} such that the proposed scheme is decodable. 
\end{enumerate}

\paragraph*{Step 1}
We choose the values for ${\bf S}_{\msf+\usf}$ with the following form,
\begin{align}
{\bf S}_{\msf+\usf}&= \left[ \begin{array}{cccccccccccc}
 b^{1,1}_{1}  & \cdots &  b^{1,1}_{\frac{\vsf}{\Ksf_{\rm c}}}  & b^{1,1}_{\frac{\vsf}{\Ksf_{\rm c}}+1} &\cdots & b^{1,1}_{\frac{2\vsf}{\Ksf_{\rm c}}}  &  b^{1,1}_{\frac{2\vsf}{\Ksf_{\rm c}}+1}  & \cdots& b^{1,1}_{\frac{(\Ksf_{\rm c}-1)\vsf}{\Ksf_{\rm c}}}  & b^{1,1}_{\frac{(\Ksf_{\rm c}-1)\vsf}{\Ksf_{\rm c}}+1}& \cdots& b^{1,1}_{\vsf} \\ 
  b^{1,2}_{1}  & \cdots &  b^{1,2}_{\frac{\vsf}{\Ksf_{\rm c}}}  & b^{1,2}_{\frac{\vsf}{\Ksf_{\rm c}}+1} &\cdots & b^{1,2}_{\frac{2\vsf}{\Ksf_{\rm c}}} & b^{1,2}_{\frac{2\vsf}{\Ksf_{\rm c}}+1}  & \cdots & b^{1,2}_{\frac{(\Ksf_{\rm c}-1)\vsf}{\Ksf_{\rm c}}}  & b^{1,2}_{\frac{(\Ksf_{\rm c}-1)\vsf}{\Ksf_{\rm c}}+1}& \cdots& b^{1,2}_{\vsf} \\ 
 \vdots &   \ddots&  \vdots&  \vdots&  \ddots&  \vdots& \vdots& \ddots&  \vdots& \vdots &\ddots &\vdots\\
 b^{1,\Ksf_{\rm c}}_{1} &\cdots &  b^{1,\Ksf_{\rm c}}_{\frac{\vsf}{\Ksf_{\rm c}}}  & b^{1,\Ksf_{\rm c}}_{\frac{\vsf}{\Ksf_{\rm c}}+1} &\cdots & b^{1,\Ksf_{\rm c}}_{\frac{2\vsf}{\Ksf_{\rm c}}} &b^{1,\Ksf_{\rm c}}_{\frac{2\vsf}{\Ksf_{\rm c}}+1} & \cdots &b^{1,\Ksf_{\rm c}}_{\frac{(\Ksf_{\rm c}-1)\vsf}{\Ksf_{\rm c}}} & b^{1,\Ksf_{\rm c}}_{\frac{(\Ksf_{\rm c}-1)\vsf}{\Ksf_{\rm c}}+1}& \cdots & b^{1,\Ksf_{\rm c}}_{\vsf} \\ 
    b^{2,1}_{1}   &\cdots &  b^{2,1}_{\frac{\vsf}{\Ksf_{\rm c}}}  & b^{2,1}_{\frac{\vsf}{\Ksf_{\rm c}}+1} &\cdots & b^{2,1}_{\frac{2\vsf}{\Ksf_{\rm c}}} &b^{2,1}_{\frac{2\vsf}{\Ksf_{\rm c}}+1}   & \cdots&  b^{2,1}_{\frac{(\Ksf_{\rm c}-1)\vsf}{\Ksf_{\rm c}}}&  b^{2,1}_{\frac{(\Ksf_{\rm c}-1)\vsf}{\Ksf_{\rm c}}+1}& \cdots & b^{2,1}_{\vsf} \\  
 \vdots &   \ddots&  \vdots&  \vdots&  \ddots&  \vdots& \vdots& \ddots&  \vdots& \vdots &\ddots &\vdots\\
       b^{\Nsf,\Ksf_{\rm c}}_{1}  &\cdots &  b^{\Nsf,\Ksf_{\rm c}}_{\frac{\vsf}{\Ksf_{\rm c}}}  & b^{\Nsf,\Ksf_{\rm c}}_{\frac{\vsf}{\Ksf_{\rm c}}+1} &\cdots & b^{\Nsf,\Ksf_{\rm c}}_{\frac{2\vsf}{\Ksf_{\rm c}}} &b^{\Nsf,\Ksf_{\rm c}}_{\frac{2\vsf}{\Ksf_{\rm c}}+1}  & \cdots & b^{\Nsf,\Ksf_{\rm c}}_{\frac{(\Ksf_{\rm c}-1)\vsf}{\Ksf_{\rm c}}} & b^{\Nsf,\Ksf_{\rm c}}_{\frac{(\Ksf_{\rm c}-1)\vsf}{\Ksf_{\rm c}}+1}& \cdots & b^{\Nsf,\Ksf_{\rm c}}_{\vsf}   
\end{array} \right]  \nonumber \\
&= 
\left[ \begin{array}{cccccccccccc}
* & \cdots & * & 0 &\cdots & 0  &0 & \cdots& 0& 0& \cdots& 0\\ 
0 & \cdots & 0 & * &\cdots & * &0  & \cdots& 0& 0& \cdots& 0\\ 
 \vdots &   \ddots&  \vdots&  \vdots&  \ddots&  \vdots& \vdots& \ddots&  \vdots& \vdots &\ddots &\vdots\\
0 &\cdots &  0 &0 &\cdots &0  & 0& \cdots & 0 &*& \cdots & * \\ 
*   &\cdots &  *  &0 &\cdots & 0  &0& \cdots& 0&  0& \cdots &0 \\  
 \vdots &   \ddots&  \vdots&  \vdots&  \ddots&  \vdots& \vdots& \ddots&  \vdots& \vdots &\ddots &\vdots\\
     0  &\cdots & 0 & 0 &\cdots & 0 &0 & \cdots &0 & * & \cdots & *
\end{array} \right]
,\label{eq:S4 assignment}
\end{align}
where each `$*$' represents an uniformly i.i.d.   symbol on $\mathbb{F}_{\qsf}$. 
More precisely,  for the $j^{\text{th}}$ linear combination transmitted by worker $n$ where $n\in [6]$,
we choose each of $b^{n,j}_{\frac{(j-1)\vsf}{\Ksf_{\rm c}}+1},\ldots, b^{n,j}_{\frac{j \vsf}{\Ksf_{\rm c}}}$
 uniformly i.i.d. over $\mathbb{F}_{\qsf}$, while setting the other variables in this linear combination be $0$. 
The above choice on ${\bf S}_{\msf+\usf}$ will guarantee that 
the  constraints in~\eqref{eq:transmit constraint} become  linearly independent  in terms of the remaining variables to be determined  in the next step.

\paragraph*{Step 2}
 We then fix one $t\in [\msf+\usf-1]$ and one $j\in [\Ksf_{\rm c}]$; thus     the constraints in~\eqref{eq:transmit constraint} become
    \begin{subequations} 
 \begin{align}
&0= \sv^{n,j} \dv_{k+(t-1)\Ksf} = \sum_{i_1 \in [ \Ksf_{\rm c}]} f_{i_1,k} \ s^{n,j}_{(t-1)\Ksf_{\rm c}+i_1}  + \sum_{i_2 \in [\vsf]} b^{n,j}_{i_2} \  a_{i_2,(t-1)\Ksf+k} \\
&= \sum_{i_1 \in [ \Ksf_{\rm c}]}f_{i_1,k} \   s^{n,j}_{(t-1)\Ksf_{\rm c}+i_1} + \sum_{i_3 \in \left[\frac{(j-1)\vsf}{\Ksf_{\rm c}}+1: \frac{j \vsf}{\Ksf_{\rm c}} \right]} b^{n,j}_{i_3} \  a_{i_3,(t-1)\Ksf+k},  \ \forall n\in [\Nsf],   k\in \overline{  \Zc_{n}}. \label{eq:step 2 constraints}
 \end{align}
   \end{subequations} 
Notice that in~\eqref{eq:step 2 constraints} the coefficients $f_{i_1,k}$ are the elements in the   demand matrix ${\bf F}$  and $b^{n,j}_{i_3} $ have been already determined in Step 1.  Hence, the constraints~\eqref{eq:step 2 constraints} are linear in terms of the variables 
\begin{align}
s^{n,j}_{(t-1)\Ksf_{\rm c}+i_1} \text{ and } a_{i_3,k_1},  \ \ \forall  n\in [\Nsf] ,   i_1 \in [ \Ksf_{\rm c}], i_3 \in \left[\frac{(j-1)\vsf}{\Ksf_{\rm c}}+1: \frac{j \vsf}{\Ksf_{\rm c}} \right],  k_1 \in [(t-1)\Ksf+1:t\Ksf]. \label{eq:variables to be solved in the step}
\end{align}
Next, we determine the values of the variables in~\eqref{eq:variables to be solved in the step} by solving linear equations. 
In~\eqref{eq:variables to be solved in the step}, there are totally 
$$
\Nsf \Ksf_{\rm c}+ \frac{\vsf}{\Ksf_{\rm c}} \Ksf = \Nsf \frac{\Ksf}{\Nsf} \usf+   (\Nsf_{\rm r}-\msf-\usf+1) \Ksf= \Ksf(\Nsf_{\rm r}-\msf +1)
$$
variables while in~\eqref{eq:step 2 constraints} there are totally
$$
\Nsf \frac{\Ksf}{\Nsf} (\Nsf_{\rm r}-\msf)= \Ksf(\Nsf_{\rm r}-\msf) 
$$
constraints. In order to determine all the variables in~\eqref{eq:variables to be solved in the step} while satisfying the constraints in~\eqref{eq:step 2 constraints}, for each $n\in [\Nsf]$,  we first choose each of 
\begin{align}
s^{n,j}_{(t-1)\Ksf_{\rm c}+(i-1)\usf +\text{Mod}(n,\usf)}, \ \forall i \in \left[\Ksf/\Nsf\right], \label{eq:choose value to solve eq}
\end{align}
uniformly i.i.d. over $\mathbb{F}_{\qsf}$.  Hence, among all the $ \Ksf(\Nsf_{\rm r}-\msf +1)$ variables in~\eqref{eq:variables to be solved in the step}, we have determined $\Nsf \frac{\Ksf}{\Nsf}=\Ksf$ variables. Thus there are $\Ksf(\Nsf_{\rm r}-\msf)$ variables to be solved by $\Ksf(\Nsf_{\rm r}-\msf)$ linear equations in~\eqref{eq:step 2 constraints}.  
It will  be proved in Appendix~\ref{sec:proof of encoding and decoding lemma} that with high probability,  these $\Ksf(\Nsf_{\rm r}-\msf)$ linear equations are linearly independent over these remaining  $\Ksf(\Nsf_{\rm r}-\msf)$ variables.  As a result, 
we have determined all the variables in~\eqref{eq:variables to be solved in the step}.

 By considering all the pairs $(t,j)$ where  $t\in [\msf+\usf-1]$ and $j\in [\Ksf_{\rm c}]$, we can determine all the elements in  ${\bf S}$  and  ${\bf F^{\prime}}$.

\paragraph*{Step 3}
  It will be proved in Appendix~\ref{sec:proof of encoding and decoding lemma} that the constraints in~\eqref{eq:decodability constraint} hold with high probability. 
Hence, we let each worker $n$ compute and send  $\Ksf_{\rm c}$ linear combinations,\\ i.e., $\sv^{n,j} {\bf F^{\prime}} [W_{1,1};\ldots;W_{\Ksf,1};W_{1,2};\ldots; W_{\Ksf,\msf+\tsf-1}]$  where $j\in [\Ksf_{\rm c}]$.

\paragraph*{Decoding phase}
Assume that the set of responding workers is $\Ac$ where  $\Ac\subseteq [\Ksf]$ where $|\Ac|=\Nsf_{\rm r}$. The master receives 
\begin{align}
{\bf X}_{\Ac}=
[\sv^{\Ac(1),1};\ldots;\sv^{\Ac(1),\Ksf_{\rm c}};\sv^{\Ac(2),1};\ldots;\sv^{\Ac(\Nsf_{\rm r}),\Ksf_{\rm c}}] \ 
\ {\bf F^{\prime}}  \ 
[W_{1,1};\ldots;W_{\Ksf,1};W_{1,2};\ldots;W_{\Ksf,\msf+\usf-1}] .
\end{align}
Since $[\sv^{\Ac(1),1};\ldots;\sv^{\Ac(1),\Ksf_{\rm c}};\sv^{\Ac(2),1};\ldots;\sv^{\Ac(\Nsf_{\rm r}),\Ksf_{\rm c}}]$ is full-rank, the master then computes
$$
[\sv^{\Ac(1),1};\ldots;\sv^{\Ac(1),\Ksf_{\rm c}};\sv^{\Ac(2),1};\ldots;\sv^{\Ac(\Nsf_{\rm r}),\Ksf_{\rm c}}]^{-1} {\bf X}_{\Ac}
$$
to obtain ${\bf F^{\prime}}  [W_{1,1};\ldots;W_{\Ksf,1};W_{1,2};\ldots;W_{\Ksf,\msf+\usf-1}]$, which contains its demanded linear combinations.

 \paragraph*{Performance}
Since each worker sends  $\frac{\Ksf_{\rm c} \Lsf}{\msf+\usf-1}$ symbols,
the   communication cost is $\frac{\Nsf_{\rm r} \Ksf_{\rm c} \Lsf}{(\msf+\usf-1)\Lsf}=\frac{\Nsf_{\rm r} \Ksf_{\rm c} }{\msf+\usf-1}$, coinciding with~\eqref{eq:achie case 1}.

\begin{rem}
\label{rem:intuition of achie constraint}
The proposed scheme works for the case where 
\begin{align}
\Nsf \geq \frac{\msf+\usf-1}{\usf}+ \usf(\Nsf_{\rm r}-\msf-\usf+1), \label{eq:work constraint}
\end{align}
 which can be explained intuitively in the following way.
It will be proved in Appendix~\ref{sec:proof of encoding and decoding lemma} that if the proposed scheme works for the 
 $\big(  \Nsf,\Nsf,\Nsf_{\rm r},  \usf , \msf \big)$ distributed linearly separable computation problem (i.e., the number of messages is equal to $\Nsf$)  with high probability, 
  then  with high probability the proposed scheme also  works for the  $\left(  \Ksf,\Nsf,\Nsf_{\rm r},  \frac{\Ksf}{\Nsf}\usf , \msf \right)$ distributed linearly separable computation problem  where    $\Nsf$ divides  $\Ksf$. 
 Hence, let us then analyse the case $\Ksf=\Nsf$.

We fix one  $t\in [\msf+\usf-1]$ in  the constraints~\eqref{eq:transmit constraint}. In Step 2 of the computing phase,   we should solve the following problem:
\paragraph*{Problem $t$}
Determine the values of   the variables 
\begin{align}
s^{n,j}_{(t-1)\usf+i_1} \text{ and } a_{i_3,k}, \ \forall n \in [\Nsf], j\in [\usf], i_1\in [\usf], i_3 \in \left[\vsf \right],  k\in [(t-1)\Ksf:t\Ksf] \label{eq:big variables} 
\end{align} 
satisfying the constraints 
\begin{align}
\sum_{i_1 \in [ \usf]}f_{i_1,k} \  s^{n,j}_{(t-1)\usf+i_1} + \sum_{i_3 \in \left[\frac{(j-1)\vsf}{\usf}+1: \frac{j \vsf}{\usf} \right]} b^{n,j}_{i_3} \  a_{i_3,(t-1)\Ksf+k}=0,  \ \forall j\in [\usf],  n\in [\Nsf],   k\in \overline{  \Zc_{n}}.
\end{align}

Notice that by solving Problem $t$, for each  $i \in [\vsf]$, we can determine 
$$[s^{1,1}_{(t-1)\usf+i};\ldots;s^{1,\usf}_{(t-1)\usf+i};s^{2,1}_{(t-1)\usf+i};\ldots;s^{\Nsf,\usf}_{(t-1)\usf+i}],$$  which is 
the $\left((t-1)\usf+i \right)^{\text{th}}$ column of ${\bf S}$.
Another important observation is that, Problem $t_1$ is totally equivalent to Problem $t_2$ for any $t_1 \neq t_2$.
  Thus, we can introduce the following unified problem. 
\paragraph*{Unified Problem}
Determine the values of   the variables 
\begin{align}
p^{n,j}_{i_1} \text{ and } q_{i_3,k}, \ \forall n \in [\Nsf], j\in [\usf], i_1\in [\usf], i_3 \in \left[\vsf \right],  k\in [\Ksf] \label{eq:unified big variables} 
\end{align} 
satisfying the constraints 
\begin{align}
\sum_{i_1 \in [\usf]}f_{i_1,k} \  p^{n,j}_{i_1} + \sum_{i_3 \in \left[\frac{(j-1)\vsf}{\usf}+1: \frac{j \vsf}{\usf} \right]} b^{n,j}_{i_3} \  q_{i_3,k}=0,  \ \forall j\in [\usf],  n\in [\Nsf],   k\in \overline{  \Zc_{n}}.
\label{eq:unified constraint} 
\end{align}

In the unified problem, there are 
$$
\Nsf \usf \usf + \vsf \Ksf= \Nsf \usf (\usf+ \Nsf_{\rm r}-\msf-\usf+1)= \Nsf \usf (  \Nsf_{\rm r}-\msf +1)
$$
variables and 
$
\Nsf\usf (\Nsf_{\rm r}-\msf)
$
constraints. 
Hence, the number of linearly independent solutions of the unified problem is no less than $ \Nsf \usf (  \Nsf_{\rm r}-\msf +1)-\Nsf\usf (\Nsf_{\rm r}-\msf)=\Nsf\usf$, where the equality holds when the constraints in the unified problem is linearly independent. To guarantee  that all the columns in ${\bf S}$ are linearly independent, we should assign $\msf+\usf-1$ linearly independent solutions  to Problems $1,2,\ldots,\msf+\usf-1$. 

In addition,  among all the linearly independent solutions of the unified problem, there are $\usf \vsf$ trivial solutions which  we cannot pick. More precisely,
for each $i  \in \left[\vsf \right]$ and $d\in [\usf]$, one possible solution is to set (recall that $\fv_d$ represents the $d^{\text{th}}$ demand vector)
$$
(q_{i ,1},q_{i ,2},\ldots,q_{i_3,\Ksf})= \fv_{d},
$$
 while setting $q_{i_3,k}=0$ if $i_3 \neq i$. 
In addition, we set 
$$
p^{n,j}_{i}= - b^{n,j}_{i} , \ \forall n\in [\Nsf], j\in[\usf],
$$
while setting $p^{n,j}_{i_1}=0$ if $i_1 \neq i$.
It can be easily checked that by the above choice of variables, the constraints in~\eqref{eq:unified constraint} holds. Hence, the above choice is one possible solution of the unified problem. There are totally $\usf \vsf$ such possible solutions.
However, any combination of such  $\usf \vsf$ solutions cannot be chosen as a solution of  Problem $t$. This is because in each of the  above solutions, 
there is a column  of ${\bf S}$ (i.e., $[p^{1,1}_{i};\ldots;p^{1,\usf}_{i};p^{1,2}_{i};\ldots;p^{\Nsf,\usf}_{i}]$), which can be expressed by  a fixed column of ${\bf S}$ (i.e., $[b^{1,1}_{i};\ldots;b^{1,\usf}_{i};b^{1,2}_{i};\ldots;b^{\Nsf,\usf}_{i}]$). Hence, the full-rank constraints in~\eqref{eq:decodability constraint} cannot hold. 

As a result, if we have 
\begin{align}
\Nsf\usf \geq \msf+\usf-1+ \usf\vsf = \msf+\usf-1+ \usf^2(\Nsf_{\rm r}-\msf-\usf+1)
\end{align}
which is equivalent to~\eqref{eq:work constraint}, it can be guaranteed that we can assign one linearly independent non-trivial solution to each Problem $t$.

 \hfill $\square$ 
\end{rem} 

 For each $\frac{\Ksf}{\Nsf}(\usf-1) < \Ksf_{\rm c} < \frac{\Ksf}{\Nsf}\usf$ where $\usf\in [\Nsf_{\rm r}-\msf+1]$, 
we first generate $\frac{\Ksf}{\Nsf} \usf-\Ksf_{\rm c}$ demand vectors whose elements are uniformly i.i.d. over $\mathbb{F}_{\qsf}$, and add these vectors into the demand matrix ${\bf F}$. Next, we use   the above distributed computing scheme with $\Ksf^{\prime}_{\rm c}= \frac{\Ksf}{\Nsf} \usf$.    
Hence, the   communication cost is     $ \frac{\Nsf_{\rm r} \Ksf^{\prime}_{\rm c} }{\msf+\usf-1}= \frac{\Nsf_{\rm r} \Ksf \usf}{\Nsf(\msf+\usf-1)}$, coinciding with~\eqref{eq:achie case 1}.

 \section{Conclusions and Future Research Directions}
\label{sec:conclusion}
In this paper, we studied the computation-communication costs tradeoff for the distributed linearly separable computation problem. 
A converse bound under the constraint of cyclic assignment was proposed, and we also proposed  a novel distributed computing scheme under some parameter regimes.  Some exact optimality results were derived with or without the constraint of cyclic assignment. 
 The proposed computing scheme was also proved to be generally order optimal  within a factor of $2$  under the constraint of cyclic assignment.

The simplest open which the proposed scheme cannot work is the case where $\Ksf=\Nsf=\Nsf_{\rm r}=5$, $\Ksf_{\rm c}=2$, and $\msf=2$. Further works include  the design of the distributed computing scheme for the  open cases  and the derivation of the converse bound for any dataset assignment.

  \appendices

\section{Feasibility Proof   of  the Proposed Computing Scheme in Section~\ref{sec:achie}}
 \label{sec:proof of encoding and decoding lemma}
 In the following, we first show that for the  $(\Ksf,\Nsf,\Nsf_{\rm r}, \Ksf_{\rm c}, \msf)=\big(  \Nsf,\Nsf,\Nsf_{\rm r},  \usf , \msf \big)$ distributed linearly separable computation problem,  where $\Nsf \geq \frac{\msf+\usf-1}{\usf}+ \usf(\Nsf_{\rm r}-\msf-\usf+1)$, the proposed computing scheme works with high probability. Next we show that 
  if the proposed scheme works for the 
 $\big(  \Nsf,\Nsf,\Nsf_{\rm r},  \usf , \msf \big)$ distributed linearly separable computation problem   with high probability, 
  then  with high probability the proposed scheme also  works for the  $\left(  \Ksf,\Nsf,\Nsf_{\rm r},  \frac{\Ksf}{\Nsf}\usf , \msf \right)$ distributed linearly separable computation problem, where $\frac{\Ksf}{\Nsf}$ is a  positive integer.
  
  \subsection{$\Ksf=\Nsf$}
  \label{sub:encoding proof N=K}
  The feasibility of the proposed computing scheme is proved by  the Schwartz-Zippel Lemma~\cite{Schwartz,Zippel,Demillo_Lipton} as we used in~\cite[Appendix C]{linearcomput2020wan} for 
the  computing scheme where $\msf=1$. For the sake of simplicity, in the following we provide the sketch of the feasibility proof.   
   
 Recall that  in Step 2 of the proposed computing scheme, for each pair $(t,j)$ where $t \in [\msf+\usf-1]$ and $j\in [\usf]$,  we need to determine the values of the variables in~\eqref{eq:variables to be solved in the step} while satisfying the linear constraints in~\eqref{eq:step 2 constraints}. In addition, 
  among all the variables in~\eqref{eq:variables to be solved in the step},  we choose   the values of the  variables in~\eqref{eq:choose value to solve eq} uniformly i.i.d. over $\mathbb{F}_{\qsf}$. 
Then there are remaining   $\Ksf(\Nsf_{\rm r}-\msf)$ variables (the vector of these $\Ksf(\Nsf_{\rm r}-\msf)$ variables is assumed to be  $\bv$) and $\Ksf(\Nsf_{\rm r}-\msf)$ linear equations over these variables, and thus we can express these linear equations as (recall that $(\mathbf{M})_{m \times n}$ indicates that the dimension of matrix $\mathbf{M}$ is $m \times n$)
\begin{align}
({\bf A})_{\Ksf(\Nsf_{\rm r}-\msf) \times \Ksf(\Nsf_{\rm r}-\msf)} \ (\bv)_{\Ksf(\Nsf_{\rm r}-\msf) \times 1} = (\cv)_{\Ksf(\Nsf_{\rm r}-\msf) \times 1}, \label{eq:before inverse}
\end{align}
where the coefficients in ${\bf A}$ and $\cv$ are composed of the elements in ${\bf F}$,  ${\bf S}_{\msf+\usf}$, and the   variables in~\eqref{eq:choose value to solve eq}  which are all generated uniformly i.i.d. over $\mathbb{F}_{\qsf}$. 
Hence, the determinant of ${\bf A}$ can be seen as a multivariate polynomial of the elements in ${\bf F}$, ${\bf S}_{\msf+\usf}$  and the   variables in~\eqref{eq:choose value to solve eq}. Since we assume $\qsf\to \infty$, by the  Schwartz-Zippel Lemma~\cite{Schwartz,Zippel,Demillo_Lipton}, if this polynomial is a non-zero multivariate polynomial  (i.e., a multivariate polynomial whose coefficients are not all $0$), the probability that the polynomial is equal to $0$ over all possible realization of  ${\bf F}$,  ${\bf S}_{\msf+\usf}$, and the   variables in~\eqref{eq:choose value to solve eq}, goes to $0$. In other words, the determinant is non-zero with high probability. 
So the next step is to show this polynomial is non-zero. This means that  we need to find one realization of  ${\bf F}$,  ${\bf S}_{\msf+\usf}$, and the   variables in~\eqref{eq:choose value to solve eq}, such that this polynomial is not equal to zero. 
By random generation of ${\bf F}$,  ${\bf S}_{\msf+\usf}$, and the   variables in~\eqref{eq:choose value to solve eq}, we have tested all cases where 
$\Nsf =\Ksf \leq 40$  satisfying the constraint $\Nsf \geq \frac{\msf+\usf-1}{\usf}+ \usf(\Nsf_{\rm r}-\msf-\usf+1)$. 
Hence, for each    pair $(t,j)$, the probability that Step 2  of the proposed computing scheme is feasible goes to $1$. By the probability union bound,  the probability that Step 2  of the proposed computing scheme is feasible for all pairs of $(t,j)$, also goes to $1$.
Moreover,   by using the the Cramer's rule, each element in $\bv$ can be seen as a ratio of two polynomials of the elements in ${\bf F}$, ${\bf S}_{\msf+\usf}$  and the   variables in~\eqref{eq:choose value to solve eq}, where the polynomial in the denominator is non-zero with high probability. 
 As a result, each element in ${\bf S}$ can be seen as  ratio of two polynomials of the elements in ${\bf F}$, ${\bf S}_{\msf+\usf}$  and the   variables in~\eqref{eq:choose value to solve eq} for all pairs $(t,j)$. So for each $\Ac \subseteq [\Nsf]$ where $|\Ac|=\Nsf_{\rm r}$, the determinant of the matrix $[\sv^{\Ac(1),1};\ldots;\sv^{\Ac(1),\Ksf_{\rm c}};\sv^{\Ac(2),1};\ldots;\sv^{\Ac(\Nsf_{\rm r}),\Ksf_{\rm c}}]$ 
can be expressed as 
$$
Y_{\Ac}= \sum_{i \in [(\Nsf_{\rm r}\usf)!]} \frac{P_i}{Q_i},
$$
where $P_i$ and $Q_i$ are polynomial of the elements in   ${\bf F}$, ${\bf S}_{\msf+\usf}$  and the   variables in~\eqref{eq:choose value to solve eq} for all pairs $(t,j)$. 
We want to prove that  $Y_{\Ac} \prod_{i \in [(\Nsf_{\rm r}\usf)!]} Q_i$ is  a non-zero polynomial such that we can use    the  Schwartz-Zippel Lemma~\cite{Schwartz,Zippel,Demillo_Lipton}
to show that the determinant $Y_{\Ac} $ is not equal to zero with high probability. 
Again, by random generation of ${\bf F}$,  ${\bf S}_{\msf+\usf}$, and the variables in~\eqref{eq:choose value to solve eq} for all pairs $(t,j)$, we have tested all cases where 
$\Nsf =\Ksf \leq 40$  satisfying the constraint $\Nsf \geq \frac{\msf+\usf-1}{\usf}+ \usf(\Nsf_{\rm r}-\msf-\usf+1)$. In these cases, with  the random choices, both $\prod_{i \in [(\Nsf_{\rm r}\usf)!]} Q_i$ and   $Y_{\Ac}$  are not equal to zero, and thus $Y_{\Ac} \prod_{i \in [(\Nsf_{\rm r}\usf)!]} Q_i$ is not equal to $0$.

In conclusion, we prove the feasibility of the proposed computing scheme in Steps 2 and 3 with high probability, for the case where $\frac{\msf+\usf-1}{\usf}+ \usf(\Nsf_{\rm r}-\msf-\usf+1) \leq \Ksf=\Nsf \leq 40$.

\subsection{$\Nsf$ divides $\Ksf$} 
\label{sub:encoding proof N divides K}
We  then consider the 
 $(\Ksf,\Nsf,\Nsf_{\rm r}, \Ksf_{\rm c}, \msf)=\left(  \Ksf,\Nsf,\Nsf_{\rm r},  \frac{\Ksf}{\Nsf}\usf , \msf \right)$ distributed linearly separable computation problem, where $\Nsf \geq \frac{\msf+\usf-1}{\usf}+ \usf(\Nsf_{\rm r}-\msf-\usf+1)$ and $\frac{\Ksf}{\Nsf}$ is a positive integer.
 Similar to the proof for the case where $\Ksf=\Nsf$,  we also aim to find a specific realization of  ${\bf F}$, ${\bf S}_{\msf+\usf}$  and the   variables in~\eqref{eq:choose value to solve eq} for all pairs $(t,j)$, such that Steps 2 and 3 of the proposed scheme are feasible (i.e., the determinant polynomials are non-zero). 
 
 We construct the demand matrix  (i.e., ${\bf F}$ with dimension $\frac{\Ksf}{\Nsf}\usf \times \Ksf$) as follows,
  \begin{align}
 {\bf F}  = \left[\begin{array}{c:c:c:c}
 ({\bf F}_1)_{\usf \times \Nsf}  & {\bf 0}_{\usf \times \Nsf}  & \cdots & {\bf 0}_{\usf \times \Nsf}   \\ \hdashline
{\bf 0}_{\usf \times \Nsf} &  ({\bf F}_2)_{\usf \times \Nsf}   & \cdots & {\bf 0}_{\usf \times \Nsf}   \\ \hdashline 
 \vdots   & \vdots  &  \vdots& \vdots \\ \hdashline
 {\bf 0}_{\usf \times \Nsf} &   {\bf 0}_{\usf \times \Nsf}    & \cdots &  ({\bf F}_{\Ksf/\Nsf})_{\usf \times \Nsf} 
 \end{array}
\right], \label{eq:constructed F for N divides K}
 \end{align} 
 where each element in ${\bf F}_i, i\in \left[\frac{\Ksf}{\Nsf} \right]$ is generated uniformly i.i.d.        over  $\mathbb{F}_{\qsf}$. 
    In the above construction, 
the   $\left(  \Ksf,\Nsf,\Nsf_{\rm r},  \frac{\Ksf}{\Nsf}\usf , \msf \right)$ distributed linearly separable computation problem is divided into $\frac{\Ksf}{\Nsf}$ independent/disjoint  $\big(  \Nsf,\Nsf,\Nsf_{\rm r},  \usf , \msf \big)$  distributed linearly separable computation sub-problems.
Since the determinant polynomials are non-zero with high probability  for each sub-problem as we proved in Appendix~\ref{sub:encoding proof N=K}, it can be seen that the    determinant polynomials for   the   $(\left(  \Ksf,\Nsf,\Nsf_{\rm r},  \frac{\Ksf}{\Nsf}\usf , \msf \right)$ distributed linearly separable computation problem
are also non-zero with high probability.

\bibliographystyle{IEEEtran}
\bibliography{IEEEabrv,IEEEexample}

\begin{thebibliography}{10}
\providecommand{\url}[1]{#1}
\csname url@samestyle\endcsname
\providecommand{\newblock}{\relax}
\providecommand{\bibinfo}[2]{#2}
\providecommand{\BIBentrySTDinterwordspacing}{\spaceskip=0pt\relax}
\providecommand{\BIBentryALTinterwordstretchfactor}{4}
\providecommand{\BIBentryALTinterwordspacing}{\spaceskip=\fontdimen2\font plus
\BIBentryALTinterwordstretchfactor\fontdimen3\font minus
  \fontdimen4\font\relax}
\providecommand{\BIBforeignlanguage}[2]{{%
\expandafter\ifx\csname l@#1\endcsname\relax
\typeout{** WARNING: IEEEtran.bst: No hyphenation pattern has been}%
\typeout{** loaded for the language `#1'. Using the pattern for}%
\typeout{** the default language instead.}%
\else
\language=\csname l@#1\endcsname
\fi
#2}}
\providecommand{\BIBdecl}{\relax}
\BIBdecl

\bibitem{amazon2015amazon}
E.~Amazon, ``Amazon web services,'' \emph{Available in: http://aws. amazon.
  com/es/ec2/(November 2012)}, 2015.

\bibitem{gcp2015}
K.~S.~P.~T. and L.~U. Gonzalez, \emph{Building Your Next Big Thing with Google
  Cloud Platform: A Guide for Developers and Enterprise Architects}.\hskip 1em
  plus 0.5em minus 0.4em\relax Apress, 2015.

\bibitem{wilder2012cloud}
B.~Wilder, \emph{Cloud architecture patterns: using microsoft azure}.\hskip 1em
  plus 0.5em minus 0.4em\relax " O'Reilly Media, Inc.", 2012.

\bibitem{speedup2018Lee}
K.~Lee, M.~Lam, R.~Pedarsani, D.~Papailiopoulos, and K.~Ramchandran, ``Speeding
  up distributed machine learning using codes,'' \emph{IEEE Trans. Inf.
  Theory}, vol.~64, no.~3, Mar. 2018.

\bibitem{linearcomput2020wan}
K.~Wan, H.~Sun, M.~Ji, and G.~Caire, ``Distributed linearly separable
  computation,'' \emph{available at arXiv:2007.00345}, Jul. 2020.

\bibitem{highdimensional2017lee}
K.~Lee, C.~Suh, and K.~Ramchandran, ``High-dimensional coded matrix
  multiplication,'' \emph{in IEEE International Symposium on Information Theory
  (ISIT)}, Jun. 2017.

\bibitem{sparsematrix2018wang}
S.~Wang, J.~Liu, , and N.~Shroff, ``Coded sparse matrix multiplication,''
  \emph{in Proc. 35th Intl. Conf. on Mach. Learning (ICML)}, pp. 5139--5147,
  2018.

\bibitem{polyoptYu2017}
Q.~Yu, M.~A. Maddah-Ali, and A.~S. Avestimehr, ``Polynomial codes: an optimal
  design for high-dimensional coded matrix multiplication,'' \emph{in Advances
  in Neural Information Processing Systems (NIPS)}, pp. 4406--4416, 2017.

\bibitem{yu2020stragglermitigation}
------, ``Straggler mitigation in distributed matrix multiplication:
  Fundamental limits and optimal coding,'' \emph{IEEE Trans. Infor. Theory},
  vol.~66, no.~3, pp. 1920--1933, Mar. 2020.

\bibitem{dutta2020optimalRT}
S.~Dutta, M.~Fahim, F.~Haddadpour, H.~Jeong, V.~Cadambe, and P.~Grover, ``On
  the optimal recovery threshold of coded matrix multiplication,'' \emph{IEEE
  Trans. Infor. Theory}, vol.~66, no.~1, pp. 278--301, Jan. 2020.

\bibitem{struresis2020ramamoo}
A.~Ramamoorthy, A.~B. Das, and L.~Tang, ``Straggler-resistant distributed
  matrix computation via coding theory,'' \emph{available at arXiv:2002.03515},
  Feb. 2020.

\bibitem{jia2019matrixCSA}
Z.~Jia and S.~A. Jafar, ``Cross subspace alignment codes for coded distributed
  batch computation,'' \emph{arXiv:1909.13873}, Sep. 2019.

\bibitem{gradiencoding}
R.~Tandon, Q.~Lei, A.~G. Dimakis, and N.~Karampatziakis, ``Gradient coding:
  Avoiding stragglers in distributed learning,'' \emph{in Advances in Neural
  Information Processing Systems (NIPS)}, p. 3368–3376, 2017.

\bibitem{MDSGC2018raviv}
N.~Raviv, R.~Tandon, A.~Dimakis, and I.~Tamo, ``Gradient coding from cyclic mds
  codes and expander graphs,'' \emph{in Proc. Int. Conf. on Machine Learning
  (ICML)}, pp. 4302--4310, Jul. 2018.

\bibitem{improvedGC2017halbawi}
W.~Halbawi, N.~Azizan-Ruhi, F.~Salehi, and B.~Hassibi, ``Improving distributed
  gradient descent using reed-solomon codes,'' \emph{available at
  arXiv:1706.05436}, Jun. 2017.

\bibitem{efficientgradientcoding}
M.~Ye and E.~Abbe, ``Communication computation efficient gradient coding,''
  \emph{in Advances in Neural Information Processing Systems (NIPS)}, pp.
  5610--5619, 2018.

\bibitem{adaptiveGC2020}
H.~Cao, Q.~Yan, and X.~Tang, ``Adaptive gradient coding,''
  \emph{arXiv:2006.04845}, Jun. 2020.

\bibitem{shortdot2016dutta}
S.~Dutta, V.~Cadambe, and P.~Grover, ``Short-dot: Computing large linear
  transforms distributedly using coded short dot products,'' \emph{in Advances
  in Neural Information Processing Systems (NIPS)}, pp. 2100--2108, 2016.

\bibitem{yangelastic2019}
Y.~Yang, M.~Interlandi, P.~Grover, S.~Kar, S.~Amizadeh, and M.~Weimer, ``Coded
  elastic computing,'' \emph{in IEEE International Symposium on Information
  Theory (ISIT)}, pp. 2654--2658, 2019.

\bibitem{replicationcode2020}
A.~Behrouzi-Far and E.~Soljanin, ``Efficient replication for straggler
  mitigation in distributed computing,'' \emph{available at arXiv:2006.02318},
  Jun. 2020.

\bibitem{Schwartz}
J.~T. Schwartz, ``Fast probabilistic algorithms for verification of polynomial
  identities,'' \emph{Journal of the ACM (JACM)}, vol.~27, no.~4, pp. 701--717,
  1980.

\bibitem{Zippel}
R.~Zippel, ``Probabilistic algorithms for sparse polynomials,'' in
  \emph{International symposium on symbolic and algebraic manipulation}.\hskip
  1em plus 0.5em minus 0.4em\relax Springer, 1979, pp. 216--226.

\bibitem{Demillo_Lipton}
R.~A. Demillo and R.~J. Lipton, ``A probabilistic remark on algebraic program
  testing,'' \emph{Information Processing Letters}, vol.~7, no.~4, pp.
  193--195, 1978.

\end{thebibliography}

\end{document}